# Late accretion to the Moon recorded in zircon (U-Th)/He thermochronometry

***Short title:*** Late accretion to the Moon

Nigel M. Kelly[a,b], Rebecca M. Flowers[a], James R. Metcalf[a], and Stephen J. Mojzsis[a,b,c]

[a]Department of Geological Sciences, University of Colorado, 2200 Colorado Avenue, UCB 399, Boulder, CO 80309-0399, USA.
[b]Collaborative for Research in Origins (CRiO), The John Templeton Foundation – FfAME Origins Program.
[c]Institute for Geological and Geochemical Research, Research Centre for Astronomy and Earth Sciences, Hungarian Academy of Sciences, 45 Budaörsi Street, H-1112 Budapest, Hungary.

Corresponding author:
Nigel M. Kelly
Department of Geological Sciences
University of Colorado Boulder
2200 Colorado Avenue, UCB399
Boulder, CO 80309, U.S.A.
*Email*: nigel.kelly@colorado.edu

***Keywords***: Moon, bombardment, Apollo 14, zircon, (U-Th)/He thermochronometry, thermal modeling

## Abstract

We conducted zircon (U-Th)/He (ZHe) analysis of lunar impact-melt breccia 14311 with the aim of leveraging radiation damage accumulated in zircon over extended intervals to detect low-temperature or short-lived impact events that have previously eluded traditional isotopic dating techniques. Our ZHe data record a coherent date vs. effective Uranium concentration (eU) trend characterized by >3500 Ma dates from low (≤75 ppm) eU zircon grains, and *ca.* 110 Ma dates for high (>100 ppm) eU grains. A progression between these date populations is apparent for intermediate (75-100 ppm) eU grains. Thermal history modeling constrains permissible temperatures and cooling rates during and following impacts. Modeling shows that the data are most simply explained by impact events at *ca*. 3950 Ma and *ca.* 110 Ma, and limits allowable temperatures of heating events between 3950-110 Ma. Modeling of solar cycling thermal effects at the lunar surface precludes this as the explanation for the *ca.* 110 Ma ZHe dates. We propose a sample history characterized by zircon resetting during the *ca.* 3950 Ma Imbrium impact event, with subsequent heating during an impact at *ca.* 110 Ma that ejected the sample to the vicinity of its collection site. Our data show that zircon has the potential to retain $^{4}$He over immense timescales (≥3950 Myrs), thus providing a valuable new thermochronometer for probing the impact histories of lunar samples, and martian or asteroidal meteorites.

## 1. Introduction

Impacts are one of the most important physiochemical processes shaping planetary surfaces. The timing and amplitude of impacts to the inner solar system is, however, debated. This is reflected in differences in long-term cratering estimates ranging from a simple monotonic decline in late accretion impact flux since crust formation, to rapid decline to near current levels followed by little change in the last ~3000 Myr (e.g., Neukum and Ivanov, 1994; Lowe et al., 2014).

The Moon is our ultimate baseline for a record of late accretion to the inner solar system. This is a consequence of the Moon's proximity to Earth, its absence of effective crustal renewal, and availability of samples collected directly from its surface. Much has been learned from high-temperature chronometers, such as U-Pb in zircon, about the early bombardment history (e.g., lunar crust formation: Nemchin et al., 2008; Taylor et al., 2009; Borg et al., 2014) and major basin-forming impacts inferred to reflect the purported Late Heavy Bombardment, or LHB (e.g., Grange et al., 2009; Hopkins and Mojzsis, 2015; Merle et al., 2017). The subsequent evolution of the impact flux is, however, still poorly constrained. While significant contributions have been made to our understanding of the cratering record from the $^{40}$Ar-$^{39}$Ar dating technique (e.g., Fernandes et al., 2013, and references therein), such studies may provide an integrated view due to the differing susceptibilities to thermal resetting among the constituents of multi-component and multi-generation lunar impact-melt breccias (Turner, 1971; Shuster et al., 2010; Shuster and Cassata, 2015; Boehnke and Harrison, 2016; Mercer et al., 2016).

Alternatively, dating single lunar zircon grains with the (U-Th)/He thermochronometry method (Reiners et al., 2002) holds promise as a means to decipher the history of late accretion to the Moon. Closure to diffusion of $^{4}$He, a stable daughter product of U, Th and Sm alpha-decay, occurs at much lower temperatures in zircon (<210°C: Reiners et al., 2002, 2004; Guenthner et al., 2013) than closure to diffusion of Pb (>1000°C). Therefore, this tool opens the door to a rich record of less energetic thermal imprints such as those from smaller, late impacts extended over the long tail of accretion (Bottke et al., 2012). Recent advances demonstrate that radiation damage can cause zircon He retentivity to vary widely (~210 to <50°C; e.g., Guenthner et al., 2013)), therefore allowing for multiple events of contrasting energy to be recorded within a single sample. Previous work using phosphate (U-Th)/He thermochronometry on meteorite samples yielded evidence for early events such as the timing of parent body formation (Min et al., 2003), as well as shock metamorphism and ejection times from planetary surfaces (Min et al., 2004). However, ZHe has hitherto not been applied to extraterrestrial samples, nor has (U-Th)/He thermochronometry of any kind been reported on lunar rocks.

Here, we report the first results for zircon grains from Apollo 14 lunar impact-melt breccia 14311. Our goal is to exploit the effects of prolonged radiation damage accumulation (≤3950 Myrs) in lunar zircon grains to constrain a record of multiple impact events of differing peak temperature and cooling rates within a single sample. Results illustrate the power of the ZHe technique to isolate lower temperature impact events inaccessible with other routinely applied dating tools on lunar rocks, allow thermal constraints to be placed on long periods of lunar history, and if integrated with other isotope systems to reveal an impact record that corresponds to the protracted record of accretion to the Moon.

## 2. Geologic setting and sample information

### *2.1. Geologic setting of Apollo 14*

Impact-melt breccias in the vicinity of the Apollo 14 landing site were sampled from the Fra Mauro formation (Fig. 1a). This formation has been interpreted as a remnant of the ejecta blanket deposited after the impact that formed the Imbrium basin (Warner, 1972; Wilshire and Jackson, 1972; Swann et al., 1977), and includes a mix of impact melt, solid fragments from the impact target, and locally derived material reworked into the ejecta blanket (Oberbeck, 1975; Wilhelms, 1987; Stöffler et al., 1989; Stöffler and Ryder, 2001). The Apollo 14 breccias were sampled in the vicinity of Cone crater (Fig. 1b), a small (340 m wide, 75 m deep) crater estimated to have formed at *ca.* 25-40 Ma in an event that excavated part of the Fra Mauro formation (Turner et al., 1971; Swann et al., 1977).

### *2.2. Sample description – impact-melt breccia 14311*

Lunar sample 14311 was chosen for (U-Th)/He analysis because it provided one of the largest and best-characterized collection of zircon mineral separates from a single sample from the Apollo breccias. The sample was collected at station Dg, at the boundary between the continuous ejecta blanket of Cone crater and discontinuous blocky ray deposits (Swann et al., 1977; Fig. 1b). A number of other small craters that penetrate Cone crater ejecta are also located nearby (e.g., Flank crater). The sample is a melt-poor, polymict impact-melt breccia composed of >75% crystalline matrix (a pyroxene and plagioclase mosaic of 5-10 µm grains), along with mineral clasts (pyroxene, plagioclase, Fe-Ti oxides), and lithic clasts that include igneous rocks and impact breccias, which are suggestive of derivation from multiple precursors that pre-date the Imbrium impact (Carlson and Walton, 1972; Simonds et al., 1977; Swann et al., 1977). Quenched impact melt in the matrix of 14311 is in very low abundance, or absent. The dominant "equant textured" crystalline matrix, however, has been interpreted to result from solid-state recrystallization within a slowly cooling ejecta blanket at temperatures of up to ~1000°C or more (Warner, 1972).

Previous zircon U-Pb geochronology for this sample (Meyer et al., 1996; Hopkins and Mojzsis, 2015; Merle et al., 2017), documented evidence for three thermal events in the history of 14311: (1) formation of *ca.* 4330 Ma crust; (2) igneous activity or crystallization of a large impact-generated melt sheet at *ca.* 4250 Ma; and, (3) *ca.* 3950 Ma impact-shocked zircon and zircon neoblasts that crystallized from impact melt. The *ca.* 3950 Ma zircon dates correlate with U-Pb phosphate (apatite, merrilite, whitlockite) geochronology obtained for this and other Apollo 14 samples (*ca.* 3934 Ma: Nemchin et al., 2009; Snape et al., 2016; Merle et al., 2017), and probably constrain the formation of the Imbrium basin. These results are consistent with $^{40}Ar/^{39}Ar$ analyses that gave an interpreted plateau age of *ca.* 3850 Ma (Stadermann et al., 1991; cf. Boehnke and Harrison, 2016), and show that resetting of high-temperature chronometers last occurred in sample 14311 at *ca.* 3950 Ma.

### *2.3. At-surface and near-surface residence time estimates*

In comparison to the Earth, lunar surfaces are extremely old because they experience limited erosion except through impact excavation. Rocks can reside for extended intervals (hundreds of millions of years) in the *near*-surface (<5 meters; Reedy and Arnold, 1972) where they are exposed to galactic cosmic rays (GCR), or for shorter intervals (millions of years) directly *at* the lunar surface where they may be exposed to heating up to temperatures of 120°C by the Sun (e.g., Turner, 1971). We make a key distinction throughout the text between *near*-surface and *at*-surface exposure intervals, both of which are important for the interpretation of ZHe dates from lunar rocks.

The cumulative *near*-surface residence time of sample 14311, expressed in terms of exposure age, has been estimated through measurement of cosmogenic isotopes produced during interaction with GCR. These ages (*ca.* 528 Myr: $^{38}Ar$, Stadermann et al., 1991; *ca.* 661 Myr: $^{81}Kr$, Crozaz et al., 1972), are markedly older than the *near*-surface exposure times of most other breccias collected close to Cone Crater (*ca.* 25 Myr: 14306, 14053, 14321; *ca.* 97-113 Myr: 14073, 14074, 14079, 14301; 260-379 Myr: 14001, 14310, 14431, 14434; Turner et al., 1971; Crozaz et al., 1972; Stadermann et al., 1991).

The GCR *near*-surface exposure ages contrast markedly with estimates of *at*-surface residence times. Micro-crater distributions indicate a single *at*-surface exposure period and suggest that the rock was not tumbled since being exposed at the surface (Horz et al., 1972; Morrison et al., 1972). The cumulative size-frequency distributions of micro-craters on multiple exposed surfaces of 14311 yield a calculated surface residence time of 0.45 to 2 Myrs (Morrison et al., 1972), consistent with estimates from cosmic-ray particle tracks (~1 Myrs; Hart et al., 1972). Thus, available data suggest that sample 14311 was exposed on the lunar surface for ≤2 Myr.

## 3. Background on (U-Th)/He thermochronometry

Retentivity of $^4$He is dependent on the amount of radiation damage accumulated in the crystal structure (Shuster et al., 2006; Flowers et al., 2009). In zircon, moderate radiation damage initially increases $^4$He retentivity, while at yet higher radiation dosages damage zones interconnect and retentivity declines (Guenthner et al., 2013). Rates of damage accumulation are dependent on time and U+Th concentration. Hereafter defined as 'effective uranium', eU, ([U] + 0.235*[Th]), this value weights the decay of the parent isotopes for their respective α-productivity (Shuster et al., 2006). Retentivity can also depend on the time-temperature (t-T) history, because radiation damage may anneal at elevated temperatures (Nasdala et al., 2002; Yamada et al., 2007). Consequently, $^4$He retentivity can vary widely even within a single zircon population, or within single grains if zoned with respect to U+Th (e.g. Danišík et al., 2017). As such, individual grains may be variably susceptible to different degrees of resetting during the same thermal event, and the response of that population to different thermal events will change over time. Zircon, depending on relationships between eU, radiation damage accumulation, and t-T history, will experience no loss, partial loss, or complete loss of $^4$He during the same thermal event. Hence, a (U-Th)/He date measured for an individual grain, and the date-eU patterns recorded by a population of grains, are an integrated function of the total t-T history experienced by that grain population and the evolving diffusivity of He during radiation damage accumulation and/or annealing events (e.g., Flowers et al., 2009).

Recent advances in our understanding of $^{4}$He diffusion in zircon have led to the development of new kinetic models that account for the evolution of $^{4}$He retentivity during both the accumulation and annealing of radiation damage (e.g., Zircon Radiation Damage Accumulation and Annealing Model, ZRDAAM: Guenthner et al., 2013). These developments allow us to use modeling approaches (e.g., HeFTy: Ketcham, 2005) to constrain the range of potential thermal histories that may explain a ZHe dataset, although some uncertainty remains in our understanding of He retentivity and annealing in highly-damaged zircon grains (see Supplementary Files). Owing to this, the (U-Th)/He system has become a rich source of geochronological information: zircon populations with a range of eU can constrain long, complicated and multi-component time-temperature paths (e.g., Guenthner et al., 2014; Johnson et al., 2017).

A secondary consideration for ZHe data interpretation is that zircon grains from lunar breccias are commonly characterized by shock damage, which can include internal brittle or plastic deformation, and complete fragmentation. Shock features within an intact zircon can reduce the grain’s diffusion domain, therefore increasing susceptibility to He loss and possibly inducing younger ZHe dates than expected based on the entire grain size. In addition, shock deformation makes grains susceptible to fragmentation during mechanical separation prior to (U-Th)/He analysis. In this way, zircon fragments may variably capture parts of the diffusive profiles, or edges of grains depleted in $^{4}$He due to alpha-ejection from the outer ~16-20 μm of the crystal during U and Th decay. These factors, and the difficulty in predicting or characterizing these features, have the potential to introduce dispersion into a ZHe dataset.

## 4. Zircon Characterization and (U-Th)/He Thermochronology:

### *4.1. Methods*

Full details on analytical methods are provided in the Supplementary Files; a summary is provided here. Individual zircon grains were chosen from polished grain mounts of three sub-samples of breccia 14311 (14311,20, 14311,50 and 14311,60) for which U-Pb, Titanium-in-zircon thermometry, and rare earth element composition data were previously acquired by high-resolution ion microprobe (Hopkins and Mojzsis, 2015). Optical light microscope, cathodoluminescence (CL) and backscattered electron images collected in that study (e.g., Fig. 2) were used to identify those features cited above that could influence ZHe data interpretation, as well as compositional zoning such as high-U rims on low-U cores.

Thirty-two individual lunar zircon grains were analyzed for ZHe thermochronometry at the University of Colorado at Boulder, following methods described in Stanley and Flowers (2016). Selected zircon grains are representative of the three previously defined U-Pb age populations (Hopkins and Mojzsis, 2015), and provide a framework within which to interpret ZHe data (Table S1). Importantly, selected grains cover a large span of eU values (14 – 297 ppm) and accumulated radiation damage. Concentrations of U, Th and Sm (in parts per million) were calculated using volume data that were extracted from high-resolution X-ray computed tomography (HR-XCT) measurements.

All measured $^{4}He$ ($^{4}He_{TOT}$) abundances were corrected for cosmogenic $^{4}He$ ($^{4}He_{COS}$) production (Tables S2, S3). $^{4}He_{COS}$ was calculated using published production values (Leya et al., 2004), corrected for the composition of the breccia matrix (Eugster, 1988; Masarik and Reedy, 1996). Production duration was computed using the maximum exposure age of sample 14311 (*ca.* 661 Ma; Crozaz et al., 1972), which corresponds to the maximum potential for generation of $^{4}He_{COS}$ in the sample. This correction is most significant for low He grains (~25% for grains 20-4_Z8 and 60-5_Z7; Table 1), but is on average <1% for the others.

Full (U-Th)/He data are reported in Table 1. Uncertainties on individual (U–Th)/He analyses in figures, tables and text are reported at 1σ, and only include propagated analytical uncertainties for measurements of He, U, Th and Sm. Uncertainties on eU include both analytical uncertainties and the maximum deviation in density from ideal pristine zircon (4.65 g/cm$^3$) and heavily radiation damaged zircon (4.05 g/cm$^3$; Salje et al., 1999).

*4.2. Results*

Based on the irregular grain shapes and truncated zoning, the grains used in this study are all considered fragments of larger grains so no alpha-ejection corrections were applied. This approach is consistent with previous meteoritic apatite (U-Th)/He studies (e.g., Min et al., 2003, 2004). Not accounting for preservation of original alpha-ejection depleted grain boundary surfaces captured by grain fragments will lead to an underestimate of radiogenic $^{4}He$ ($^{4}He_{RAD}$). The implications of this effect for our results are discussed further below. Moreover, four grains included in this study exhibit partial rims, which based on weak CL emission suggest they may be higher in U than their respective cores. Where not correcting for alpha-ejection, high-U rims can also lead to younger ZHe dates due to a measured “excess” of U. Alternatively, where a high-U rim has been lost from a lower-U core during sample fragmentation, excess $^{4}He$ due to implantation may occur and an older apparent date measured for the grain fragment. However, the measured date-eU patterns (below) from the four zircon fragments with rims, when compared with the broadly consistent overall trend of data would suggest that the presence of high-U rims does not strongly influence the ZHe results.

Our ZHe dates range from 4587 ± 555 Ma Ma, to 6 ± 2 Ma (Fig. 3). The oldest two dates in this collection (60-5_Z1: 4587 ± 555 Ma; 20-4_Z2: 4369 ± 407 Ma) are imprecise and overlap within uncertainty with their $^{207}Pb/^{206}Pb$ ages (3960 ± 18 Ma and 3964 ± 28 Ma, respectively). The remaining ZHe data are characterized by predominantly *ca.* 4278 to 3500 Ma dates for low-eU grains (<75 ppm), a group of dates from *ca.* 212 to 6 Ma at high-eU (100 – 300 ppm), and a continuous age progression at intermediate-eU values (75-100 ppm) between the older and younger groups. The older, low-eU group of ZHe dates overlap previously published U-Pb ages for zircon and apatite from 14311 (*ca.* 3950-3940 Ma; Hopkins and Mojzsis, 2015; Snape et al., 2016; Merle et al., 2017). Of the younger group of high-eU dates, a cluster of four analyses has a weighted average age of 111 ± 26 Ma.

## 5. Discussion

### *5.1. ZHe dates from ancient lunar zircon*

The best-defined features of our ZHe data set are the *ca.* 3950-3500 Ma dates at low eU (≤75 ppm) and the *ca.* 110 Ma dates at high eU (≥100 ppm) that form part of a coherent date-eU trend (Fig. 3). The negative correlation between ZHe date and eU is a common feature in typical terrestrial zircon samples (e.g., Guenthner et al., 2013, 2014; Johnson et al., 2017), and demonstrates the effect of radiation damage on $^{4}He$ retentivity in zircon. While there is some dispersion in the data from low-eU grains, our data demonstrate that despite almost 4 billion years of radiation damage accumulation, zircon from lunar impact melt breccia sample 14311 have retained dates that exceed *ca.* 3500 Ma, with some at least as old as the major Imbrium basin-forming event at *ca.* 3950 Ma (e.g., Nemchin et al., 2009, 2017, and references therein). Dispersion in the >3500 Ma ZHe dates is discussed below.

The *ca.* 110 Ma dates suggest a thermal event that reset high-eU grains (≥100 ppm) at *ca.* 110 Ma. This group of *ca.* 110 Ma dates comports with GCR near-surface exposure ages from 4 other Apollo 14 samples (*ca.* 97 - 113 Ma: 14301, 14073, 14074, 14079; Stadermann et al., 1991, Crozaz et al., 1972). The *ca.* 110 Ma ZHe age is also compatible with the near-surface $^{38}$Ar and $^{81}$Kr exposure ages of *ca.* 528 Ma and *ca.* 661 Ma for sample 14311, despite being younger than those results. This is because diffusivity of $^{4}$He in high-eU (radiation-damaged) zircon is higher than that for bulk rock diffusion of $^{38}$Ar and $^{81}$Kr (Guenthner et al., 2013; Shuster and Cassata, 2015), and so resetting of high-eU zircon at *ca.* 110 Ma could occur with little or no resetting of cosmogenic $^{38}$Ar and $^{81}$Kr. Also, cosmogenic stable isotope accumulation requires exposure to GCRs and so record cumulative residence of a rock in the *near*-surface (<5 meters; Reedy and Arnold, 1972). Therefore, exposure ages may reflect more than one exposure event - a well-known and expected effect of impact gardening - provided that samples are not heated sufficiently for cosmogenic isotopes to be lost through diffusion. In the case of 14311, the pre-110 Ma near-surface exposure ages can be simply explained by GCR exposure prior to the thermal event recorded by *ca.* 110 Ma dates.

The thermal history modeling described next better constrains the significance of the ZHe data patterns, with specific focus on the *ca.* 110 Ma dates for high eU zircon. Given the known history of the Moon and the absence of igneous events at *ca.* 110 Ma, the two simplest possible explanations for the *ca.* 110 Ma dates are: 1) an extended period of solar heating at the lunar surface, because the exteriors of objects on Moon's surface can reach peak temperatures of 120°C during each lunar day, or 2) a *ca.* 110 Ma impact event. We quantitatively evaluate these two possibilities with thermal history modeling, and then present our preferred history for lunar impact-melt breccia 14311 consistent with our new constraints.

### *5.2. Thermal history modeling: approach and caveats*

Modeling was performed with the software HeFTy (Ketcham, 2005) and the ZRDAAM kinetic model (Guenthner et al., 2013); ZRDAAM is the only ZHe kinetic model that includes the well-documented effects of radiation damage accumulation and annealing on He diffusivity. Zircon compositions used in models (10-300 ppm) encompass the range of eU values of the grains in our lunar data set (Table 1). The grain radius chosen for modeling (100 μm) was based on lunar grain sizes following mechanical separation and takes into account the fact that analyzed grains were fragments of the original (larger) *in situ* grains. No alpha-ejection correction was applied in the models so as to provide a direct comparison to the measured data.

HeFTy diffusion models use the zircon grain radius to define the diffusion dimension, so that our model grain size represents a probable maximum in the lunar sample prior to separation. Of our analyzed grains, the modeled grains best represent fragments that originated near the center of a crystal and lacked impact shock features. Smaller grains, those with substantial impact shock features (e.g., Fig. 2c, f), or zircon grains with sub-micron defects at scales below the resolution of the imaging techniques used in this study, will have smaller effective diffusion dimensions. These grains or grain fragments would be less retentive and therefore more likely to be reset at lower temperatures, giving younger ZHe dates for the same conditions (temperature, time, zircon eU) compared to the modeled grains. Similarly, fragments originating close to the edges of large grains, which would likely have lost He through diffusion and/or alpha-ejection, would also give younger ZHe dates than predicted. These factors are potential causes for dispersion from the simplified date-eU trends generated by the models and would most commonly result in younger ZHe dates than predicted (see Brown et al., 2013). Partly for this reason, and to avoid the pitfall of over-interpretion for a complex system, we focus on the general pattern of the entire data set and on averaged dates of distinct groups of grains.

The influence that varying crystal and diffusion domain size has on the overall interpretations of the lunar data set presented here are likely minimal. First, there is no systematic relationship between grain size (measured sizes of grain fragments prior to analysis) or calculated volume (from HR-XCT data) and (U-Th)/He date (Table 1; Fig. S1). Second, there is no systematic variation between most shocked grains (e.g., those from the *ca.* 3950 Ma U-Pb zircon population) and those with no apparent shock features (Fig. 3). Third, and perhaps most important, the measured data define a coherent date-eU trend, characterized by a distinct younger group of ages (<212 Ma) that occur across a wide range of eU values (100 – 300 ppm). Two grains at the smaller end of the size spectrum clearly lie off this trend - one with distinct shock features (60-5_Z7; Fig. 2f) and another without (60-5_z9) – and both have younger ZHe dates than would be predicted for their eU (*ca.* 32 Ma and 115 Ma at eU = 48 and 54 ppm, respectively). These grains are only moderately radiation damaged (based on eU), and therefore it is probable that the diffusion domain size, and so He retentivity, were reduced by the effects of impact shock. However, we conclude from the overall data patterns that the effects of crystal and diffusion domain sizes described above do not significantly alter the larger, first-order data patterns on which our interpretations are based.

### *5.3 Evaluating solar thermal cycling effects*

The absence of an atmosphere on the Moon leads to a wide temperature variation on the lunar surface (equatorial surface temperatures may range from a maximum of ~120°C to a minimum of -180°C: Vaniman et al., 1991; Vasavada, 2012). Cyclic heating of material at the surface during each lunar day (~29.5 Earth days) has the potential to cause partial to complete diffusive loss of noble gases from crystals and glass (e.g., $^{40}Ar/^{39}Ar$: Turner, 1971; Shuster and Cassata, 2015). The latent effect of solar cycling on ZHe dates from sample 14311 zircon grains was modeled using the computer program HeFTy, incorporating the "effective diffusion temperature" (EDT) approach (Tremblay et al., 2014; Shuster and Cassata, 2015). The EDT represents the temperature corresponding to the mean diffusivity over a variable, or cyclic, temperature function (Tremblay et al., 2014), and will be greater than or equal to the mean daytime temperature (in the case of the lunar surface solar cycle), but lower than the maximum surface temperature, due to the exponential relationship of diffusivity to temperature. Many of the variables that affected the actual temperatures experienced by the grains (e.g., depth of the grains in the sample, effects of shading, direction the sample was facing) are unknown. We therefore compared maximum solar heating effects (where we assume zircon grains are fully exposed at the surface of the sample, with maximum daytime temperature = 120°C), with a condition that reflects moderate thermal attenuation due to grains being deeper within the sample (where we assume the sample was shaded for part of the lunar day or was partially covered by regolith, with peak daytime temperature = 100°C). Our HeFTy models were run for durations of heating that correspond to *at*-surface exposure (0.45 – 2 Myr; Morrison et al., 1972; Hart et al., 1972). Details on the application of the EDT technique and HeFTy modeling of solar cycling are provided in Supplementary Files (Fig. S2, S3).

These modeling results suggest that solar thermal cycling effects alone, without any impact heating post-3950 Ma, cannot explain the ZHe date-eU pattern of our lunar sample. For example, the red dashed curve in Figure 4a and 4b, which represents heating at maximum surface temperatures (120°C) and at maximum durations of exposure (2 Myrs), predicts ZHe dates much older than observed at intermediate eU (75-150 ppm) and younger than observed at higher eU (>225 ppm). The deviation of modeled values from our measured data set is more pronounced for shorter durations of exposure or where peak solar heating is reduced. To achieve resetting in low- to intermediate-eU zircon in order to generate date-eU patterns similar to our measured data, much higher temperatures than experienced on the Moon's surface, or longer periods of exposure than explained by micro-pitting studies (exceeding 660 Ma) are required. We therefore next consider impact event histories that can more plausibly account for our dataset.

### *5.4. Evaluating the thermal significance of impact events*

#### *5.4.1. Approach*

Our thermal history modeling of impact events to explain the measured ZHe date-eU pattern (Fig. 3) has two primary goals. First, because the oldest ZHe dates are the age of the Imbrium event, we aim to determine the minimum conditions necessary during the Imbrium impact to reset the (U-Th)/He date in all zircon grains to younger than the zircon crystallization age constrained by U-Pb. Second, we seek to evaluate if a post-Imbrium impact event is the most likely explanation for *ca.* 110 Ma ZHe dates. More than 1000 models covering greater than 20 thermal history scenarios were run (summarized in Table 2). Model scenarios aimed to evaluate any potential event that may have affected the zircon across the entire lunar history from *ca.* 4330 Ma.

Forward models assumed rapid heating (minutes) to peak temperatures that varied from 1500°C to 50°C, and linear cooling trajectories over durations of >10 kyr to 1 minute (time taken to cool from peak-temperature to 0°C; e.g., Fig. 5a). Parameters were selected to cover likely thermal conditions for a range of impact cratering event scenarios, because ejecta blanket dimensions, total melt volumes and clast-melt ratios will influence peak temperatures attained during thermal equilibration of solids and melt, as well as rates of cooling following emplacement (Onorato et al., 1978; Prevec and Cawthorn, 2002). For example, Figure 5a summarizes modeled event histories where zircon grains were fully reset during an impact at 3950 Ma, and then experienced a second impact event at 110 Ma. This model tests variations in cooling durations and peak temperatures, simulating a range of conditions from conductive cooling in an ejecta blanket to rapid quenching of melt-bearing rocks.

The coherent trend in the measured ZHe data provide our best constraints on valid versus invalid t-T histories. We use the following parameters to determine closeness of fit of a model output to the measured ZHe data. To evaluate the 3950 Ma event, we assessed the temperature at which all zircon grains were fully reset to 3950 Ma for a range of cooling durations (e.g., 10 kyrs). To evaluate potential impact histories and associated temperatures that could lead to the *ca.* 110 Ma ZHe dates, we consider both the generation of *ca.* 110 Ma dates in high-eU (≥100 ppm) zircon, and the preservation of >3500 Ma dates in low-eU zircon (≤75 ppm; summarized in Fig. 5b). For a thermal event history to be considered permissible, we tracked resetting of the most resistant zircon in the high-eU group (eU = 150 ppm) to where its ZHe date fell below 250 Ma, while still retaining >3500 Ma dates in 75 ppm eU zircon (the most sensitive to resetting during post-Imbrium events). All other conditions are precluded. An example of this output is summarized in Figure 5c. Modeled date-eU plots that represent key thermal event histories are presented in Figure 6, and discussed in more detail next.

*5.4.2. Implications*

Using the methodology above we can impose constraints on thermal conditions for the last *ca.* 3950 Ma of the lunar history for Apollo sample 14311, as summarized in Figure 7. First, the data constrain the minimum temperatures/cooling durations required during an impact at 3950 Ma to completely reset all of the zircon that crystallized at *ca.* 4300 Ma and *ca.* 4250 Ma, and thus replicate the *ca.* 3950-3500 Ma dates for zircon with <100 ppm eU. For cooling durations of 10 kyrs (a conservative estimate for conductive cooling in an ejecta blanket of ~350 m thickness; Abramov et al., 2013), minimum peak temperatures of only ~350°C are required to completely reset the ZHe dates (Figs. 6a, 7a, S4). Therefore, at conditions within an ejecta blanket consistent with those invoked to explain the equant matrix texture in sample 14311 (>1000°C; Warner, 1972), even relatively short event durations (e.g., 1 month, Fig. S4) would result in complete resetting of ZHe dates following deposition in an ejecta blanket at *ca.* 3950 Ma. An important additional outcome of models that solely involve heating during an impact at *ca.* 3950 Ma is that they do not replicate the measured lunar date-eU patterns from 14311; they all lack the young (*ca.* 110 Ma) dates preserved for high-eU zircon (Figs. 6a, S4).

Second, and perhaps most significantly, we find that only a thermal history including a heating event at *ca.* 110 Ma fully reproduces our observed date-eU pattern (Fig. 6b-d; compare Figs. S4-S8), even considering possible over- or under-estimates of He retentivity in highly radiation damaged grains (see Supplementary Files for detailed discussion). We also find that only a narrow range of thermal conditions are permissible during the *ca.* 110 Ma event (Fig. 7b). Therefore, if we assume a probable maximum cooling duration of 1 year (the absence of abundant quenched melt would suggest the sample cooled quickly), peak temperatures are limited to between 400 and 365°C to reset high-eU zircon to ZHe dates of *ca.* 110 Ma while still preserving >3500 Ma dates in low-eU zircon. For shorter cooling durations (e.g., 1 day), acceptable maximum temperatures are higher, but are still restricted (630-570°C). These permissible temperatures would be reduced by up to 130°C for smaller grain radii (e.g., 25 μm; Fig. S9). Thermal history models also show that the scatter in low-eU ZHe dates to as young as *ca.* 3500 Ma can be explained by a simple event history involving a single impact event at *ca.* 110 Ma overprinting zircon grains that were fully reset at *ca.* 3950 Ma. Further, it is possible that the younger 6 Ma date for the highest-eU zircon (~300 ppm) of our dataset can be explained by the superimposed effect of solar heating at lower temperatures and over shorter timescales subsequent to the 110 Ma impact event (Fig. 4a, b).

Finally, the data limit permissible maximum temperatures of any heating event that may have affected sample 14311 between *ca.* 3950 and 110 Ma by precluding thermal conditions that would reset low eU zircon to dates younger than observed (*ca.* 3500 Ma; Fig. 7c). For example, heating during basaltic volcanism at *ca.* 3400 Ma (Pidgeon et al., 2016) could not have exceeded peak temperatures of 350°C (for a cooling duration of 10 kyr; Fig. 7c; Fig. S6). If the sample was affected by the Copernicus impact (*ca.* 800 Ma; Merle et al., 2017), peak temperatures for that event could not have exceeded 450°C (for a cooling duration of 1 yr; Fig. S9).

To summarize, the ZHe data for Apollo sample 14311 are most consistent with an older (originally *ca.* 3950 Ma) zircon population that experienced a thermal event at *ca.* 110 Ma years (Fig. 6b; S6-8). The results from this single sample 1) constrain minimum temperatures and cooling durations during the *ca.* 3950 Ma (Imbrium?) impact event, 2) demonstrate the necessity for a heating event at *ca.* 110 Ma to explain the measured date-eU pattern, 3) impose tight t-T limits of this *ca.* 110 Ma heating event, and 4) limit maximum temperatures of any event between 3950 and 110 Ma.

### *5.5. Preferred model for the history of lunar impact-melt breccia 14311*

The ZHe dates and thermal history modeling results can now be integrated with our current knowledge of lunar history to expand our understanding of impact melt breccia 14311. Based on this knowledge, including the timing of post-Mare volcanism, *ca.* 110 Ma ZHe dates in sample 14311 are best explained by heating during a young impact event. The absence of abundant melt in the sample (a small volume of quenched melt has been reported; Warner, 1972), suggests that there was a low melt-solid ratio in the ejected material. This could result from a minor impact event involving a high degree of mechanical mixing of source material and a short-lived heating history. Our results limit the peak temperatures during this heating to between 400°C (1 year cooling) and 630°C (1 day cooling; Fig. 7b). It is therefore likely that sample 14311 was heated during a non-basin-forming impact event and underwent rapid cooling, with any impact melts present quenched before more extreme heating of solid fragments could occur.

The coherent date-eU pattern in the ZHe dataset, despite containing zircon with pre-Imbrium U-Pb ages and shocked zircon with ages partially to completely reset to *ca.* 3950 Ma (Hopkins and Mojzsis, 2015), suggests these grains were deposited together as part of ejecta from a major basin-forming event (possibly Imbrium) and likely share the same post-Imbrium thermal history. We favor the scenario in which ejecta materials were subsequently relocated to Fra Mauro during formation of a younger, smaller impact elsewhere. No *ca.* 110 Ma craters in the immediate vicinity of the Apollo 14 landing site are sufficiently large to cause the thermal resetting required by the ZHe data, and there is no evidence for nearby magmatism at *ca.* 110 Ma. It is probable that 14311 was sourced through ejection from a more distal location.

Tycho crater, ~86 km in diameter and located in the southern lunar highlands and dated at *ca.* 96-110 Ma from exposure ages of samples from Apollo 17 (Drozd et al., 1977; Arvidson et al., 1972), is a plausible candidate (Fig. 1a). Ejecta from this impact covers an area of up to ~560,000 km$^2$ (Fig. 1a), and discontinuous ejecta rays extend as far as Mare Serenitatis, Mare

Fecunditatis, and the vicinity of the Apollo 14 landing site (Dundas and McEwen, 2007). At ~1200 km from Tycho to the Apollo 14 sampling site, any material ejected to Fra Mauro would likely have contained low melt volumes (Onorato et al., 1978; Prevec and Cawthorn, 2002; Abramov et al., 2012) and cooled rapidly. The formation of "equant" matrix texture preserved in 14311, however, requires slow cooling at high-temperatures within a thick blanket of ejecta (Warner, 1972). It is not clear if this would be possible in the vicinity of Tycho, close to 1800 km from the edge of the Imbrium basin. Therefore, other, yet to be identified craters closer to Fra Mauro must also be considered.

An exotic origin for sample 14311 has previously been proposed. For example, contrasting U-Pb age spectra from 14311 zircon grains compared with other Apollo 14 samples suggests a source separate to the Fra Mauro formation in the Apollo 14 vicinity (Merle et al., 2017). Radiation-damage model ages (*ca.* 3400 Ma) were previously linked to mare magmatism (Pidgeon et al., 2016). Coupled with an old $^{81}$Kr exposure age (*ca.* 661 Ma; Stadermann et al., 1991), it was suggested 14311 could have been ejected from the *ca.* 800 Ma Copernicus impact crater (Pidgeon et al., 2016; Merle et al., 2017). Although the ZHe data cannot preclude heating at *ca.* 3400 Ma or involvement in the Copernicus impact event, the ZHe results do limit possible thermal conditions associated with these proposed events and require a heating event at *ca.* 110 Ma is to explain the data patterns (Fig. 7).

Based on the current dataset, we prefer a simple event scenario to explain the ZHe data characterized by impact and deposition in the Imbrium ejecta blanket at *ca.* 3950 Ma, followed by a second impact event at *ca.* 110 Ma, and limited solar heating from <2.0 Myrs. This simple scenario, which includes a non-Fra Mauro formation source for sample 14311, is consistent with the measured GCR exposure ages for the rock ($^{38}$Ar: 528, $^{81}$Kr: 661 Ma; Crozaz et al., 1972; Stadermann et al., 1991). These exposure ages likely reflect the cumulative time spent by 14311 in the *near*-surface prior to ejection to Fra Mauro combined with that spent *near*- or *at*-surface since deposition at *ca.* 110 Ma. Although 14311 was previously interpreted to belong to a suite of breccias excavated from the Fra Mauro formation by the impact that formed Cone crater (*ca.* 25 Ma; Crozaz et al., 1972; Stadermann et al., 1991), young *at*-surface ages for 14311 and another impact-melt breccia 14301 (<2 Ma; Hart et al., 1972; Horz et al., 1972; Morrison et al., 1972), and *ca.* 100 Ma GCR exposure ages for samples 14301, -073, -074, and -079, mean that this is unlikely. Instead, we argue that these samples were deposited in the Apollo 14 landing site at *ca.* 110 Ma, and sample 14311 was fragmented and moved through impact gardening in more recent times.

## 6. Conclusions

The ZHe thermochronometry of lunar Apollo samples provides a new means to document the history of late accretion to the Moon. Our (U-Th)/He data from 32 zircon grains separated from impact-melt breccia Apollo sample 14311 define a coherent date-eU trend that demonstrates the radiation damage effect on $^4$He retentivity in zircon. Forward modeling of impact and other thermal histories using the measured date-eU trends and the most recent zircon He diffusion kinetic model leads us to three major conclusions:

1) Our ZHe dates are best explained by a simple impact history characterized by a *ca.* 110 Ma thermal event, recorded in high (>100 ppm) eU zircon, which overprinted grains that were previously fully reset at *ca.* 3950 Ma during a basin-formation event, possibly Imbrium. We interpret the *ca.* 110 Ma dates to reflect resetting during a small, non-basin forming impact (Tycho?) that ejected the sample to the Apollo 14 sampling site.
2) Modeling constrains conditions during heating associated with this young impact (Fig. 7b), which requires preservation of >3500 Ma dates in low eU zircon while recording *ca.*

110 Ma dates in high-eU zircon. In addition, modeling using the EDT approach precludes solar thermal cycling on the lunar surface as the sole cause of the *ca.* 110 Ma dates.

3) While the ZHe data cannot preclude a sample history involving thermal or impact events between 3950 – 110 Ma, the models restrict maximum permissible conditions of any thermal excursions (Fig. 7c).

The results of this study also show that lunar zircon can retain $^{4}$He over immense timescales (≥3950 Myrs) and therefore provides a useful new thermochronometer for dating the impact histories of lunar rocks. While (U-Th)/He data from impact melt breccia 14311 indicates this sample experienced two major crater-forming events, we expect to obtain very different date-eU patterns from other samples of the lunar surface that experienced contrasting impact histories. Consequently, depending on degrees of overprinting, each sample may access different parts of the lunar impact history. By expanding work on the Moon to zircon occurrences documented in martian and asteroidal meteorites, the combined U-Pb and (U-Th)/He thermochronometry technique opens the door to a more comprehensive picture of long-term accretion rates in the inner solar system over a wide range of impact magnitudes.

**ACKNOWLEDGEMENTS.** We thank the brave crew of Apollo 14 (1971) who collected these samples. This study was supported by the NASA Cosmochemistry program (Grant # NNX14AG31G to SJM, RMF and JRM), by the National Science Foundation (NSF-EAR 1126991 to RMF), and by the John Templeton Foundation – FfAME Origins Program which supports the Collaborative for Research in Origins (SJM): The opinions expressed in this publication are those of the authors, and do not necessarily reflect the views of the John Templeton Foundation. SJM also acknowledges generous sabbatical leave support from the Earth Life Science Institute (ELSI) at Tokyo Institute of Technology. We also thank CAPTEM, and the JSC-Curatorial Facility directed by R. Zeigler for providing lunar sample materials for this study. E. Ellison is thanked for assistance with micro-Raman analysis. We also thank R. Ketcham, J. Maisano, M. Schindler and M. Prasad for access to HRXCT facilities and assistance with data post-processing. M. Hopkins-Wielicki is thanked for U-Pb date compilations. We are further grateful to R. Ketcham for assistance with the HeFTy software and discussion of forward modeling approaches. We also thank insightful reviews by P. Reiners and V. Fernandes, which helped to improve the clarity of the manuscript. Discussions with R. Brasser, W. Bottke, S. Marchi and O. Abramov also helped to improve this work.

## FIGURE CAPTIONS

**Fig. 1.** (a) Location of the Apollo 14 landing site within the Fra Mauro formation (shaded tan area). Image shows extent of the Imbrium basin (solid white line), location of the Tycho impact crater (red oval) with approximate extent of the Tycho continuous ejecta blanket (shaded light orange) and minimum reach of discontinuous ray deposits (dashed white line) as mapped by (Dundas and McEwen, 2007). Lunar image collected by the Lunar Rover Orbital Camera and used courtesy of NASA/GSFC/Arizona State University. (b) Map of the Apollo 14 landing site including mapped deposits of the Fra Mauro formation and ejecta associated with the *ca.* 25 Ma Cone Crater (modified from Swann et al., 1977). Location of the two Extravehicular Activity (EVA) traverses and sampling sites are given. 'LM' refers to the location of the Lunar Module. Sample 14311 was collected at site 'Dg'.

**Fig. 2.** SEM cathodoluminescence images of zircon grains from lunar sample 14311 representative of the three U-Pb populations defined by SIMS and covering a range of ZHe dates: a) zircon fragment with truncated igneous zoning; b) zircon fragment with faint zoning that may be igneous or may have formed through growth from an impact melt sheet; c) zircon grain preserving healed fractures and planar deformation features; d) and e) zircon fragments from the *ca.* 110 Ma ZHe date population; and f) zircon grain with a ZHe date that lies off the dominant date-eU trend, see text for explanation. SIMS $^{207}Pb/^{206}Pb$ ages from Hopkins and Mojzsis (2015), ZHe dates and eU values from this study. All quoted age uncertainties are $1\sigma$.

**Fig. 3.** (U-Th)/He date vs eU plot for 32 zircon grains analyzed from Apollo 14 impact-melt breccia 14311. Individual ZHe data points (error bars at $1\sigma$) are color coded according to U-Pb age population defined by Hopkins and Mojzsis (2015): blue for 4.33 Ga, green for 4.25 Ga, and orange for 3.95 Ga. Also indicated are potential events in the sample history, including a "radiation damage model age" calculated by (Pidgeon et al., 2016). P-Ne: Pre-Nectarian, Ne:

Nectarian, Im: Imbrian, Er: Eratosthenian, Co: Copernican. [a](Jacobson et al., 2014, and references therein), [b](Borg et al., 2014), [c](Hopkins and Mojzsis, 2015, Merle et al., 2014, Snape et al., 2016), [d](Pidgeon et al., 2016), [e](Stöffler and Ryder, 2001), [f](Drozd et al., 1977); Basin sequence based on (Fassett et al., 2012)

**Fig. 4.** Predicted ZHe date-eU patterns from thermal history forward modeling to simulate the effects of maximum daily heating of zircon grains on the lunar surface using the "EDT" approach (after Tremblay et al., 2014). Measured ZHe date-eU patterns are shown in gray for reference. The 120°C curves (red) represent the effects of "maximum" daytime temperature on a fully exposed surface of a sample during cycled solar heating near the lunar equator (estimated range of *at*-surface exposure periods of 0.45 and 2.0 Myrs). The 100°C curves (blue) represent attenuated heating where zircon grains were sited deeper within the sample, the sample was shaded for part of the lunar day, or the sample was partially covered by regolith. a) shows predicted date-eU patterns for a thermal history characterized by a single impact at 3950 Ma (peak temp of 1000°C and 2 kyr cooling duration) and later solar cycling, with the results plotted at different scales. The inset is zoomed to highlight high-eU data. The predicted ZHe dates do not reproduce the observed ZHe data. b) are same plots as a), but additionally include an impact event at 110 Ma (peak T = 400°C, cooling over 1 year). Inclusion of the 110 Ma heating event is required to replicate the data patterns.

**Fig. 5.** (a) Schematic representation of an example temperature-time (t-T) history used in HeFTy forward models (Ketcham, 2005). In this event series, conditions for 'Impact 1' (3950 Ma) were fixed, while peak temperatures and cooling durations for 'Impact 2' (110 Ma) were varied. (b) Lunar ZHe date-eU plot showing criteria used for determining permissible versus precluded thermal histories for a 110 Ma impact event. The predicted ZHe date-eU distribution from each forward model was compared with the observed "low-eU" (<100 ppm) and "high-eU" (>100

ppm) data encompassed by the gray boxes. Viable t-T paths are those that predict low-eU ZHe dates >3500 Ma (above horizontal red dashed line) and high-eU ZHe dates <250 Ma (below horizontal blue dashed line). These criteria result in a limited range of permissible peak temperatures for the 110 Ma event. c) ZHe date-eU patterns predicted by for the thermal history presented in (a), where all is fixed except for the peak temperature attained during the 110 Ma heating event. Using the rationale presented in (b), peak temperatures >275°C and those <250 °C are precluded. The former predicts low-eU ZHe dates younger than observed, and the latter predicts high-eU ZHe dates older than observed.

**Fig. 6.** Example predicted ZHe date-eU patterns from thermal history forward models characterized by a) zircon crystallization at 4330 Ma followed by heating during a single impact at 3950 Ma of varied peak temperature; b) heating during impact events at 3950 Ma and 110 Ma, with peak temperatures of the 110 Ma event varied and using a fixed 1 month cooling duration (that contrasts with the 2 kyr cooling duration of the forward models in Figure 5C); c) heating during an impact event at 3950 Ma followed by a heating event at 3400 Ma of variable peak temperature; d) heating during impact events at 3950 Ma and 800 Ma, with peak temperatures of the 800 Ma event varied. Only t-T histories similar to those in Figures 6B and Figure 5C, which include reheating at 110 Ma, can reproduce the data.

**Fig. 7.** Plots of peak temperature (°C) versus cooling duration following peak temperature (Kyrs) showing permissible and precluded bounds for impact events affecting breccia 14311 based on thermal history modeling results. a) Permissible thermal conditions during an impact event at 3950 Ma, required to fully reset low-eU (75 ppm eU) zircon as observed. b) Permissible thermal conditions during an impact event at 110 Ma. The black curve reflects the minimum temperatures required to reset high-eU (150 ppm eU) zircon dates to *ca.* 110 Ma, while the red curve gives the maximum allowable temperatures to still preserve >3500 Ma dates in low-eU

(75 ppm eU) zircon. c) Permissible thermal conditions for any event that may have affected sample 14311 between 3950 Ma and 110 Ma. Maximum permissible temperatures decrease with time due to accumulated radiation damage in zircon, as shown by the red and black curves.

**Fig. 8.** Preferred model for the displacement of materials now constituting impact-melt breccia 14311, consisting of a simple event scenario characterized by *ca.* 3950 Ma impact and deposition in the Imbrium ejecta blanket, a second impact event at *ca.* 110 Ma, and limited solar heating from <2.0 Myrs. Cratering schematic modified from (Osinski et al., 2011). Zone of melt production during initial impact is given by the dashed circle, and zone of vaporization shaded white circle. Preservation of *ca.* 4330 Ma and 4250 Ma U-Pb ages by some zircon grains likely requires target rocks were within the zone of mechanical mixing, while extensive resetting in grains with impact shock features may indicate location within the zone of melting.

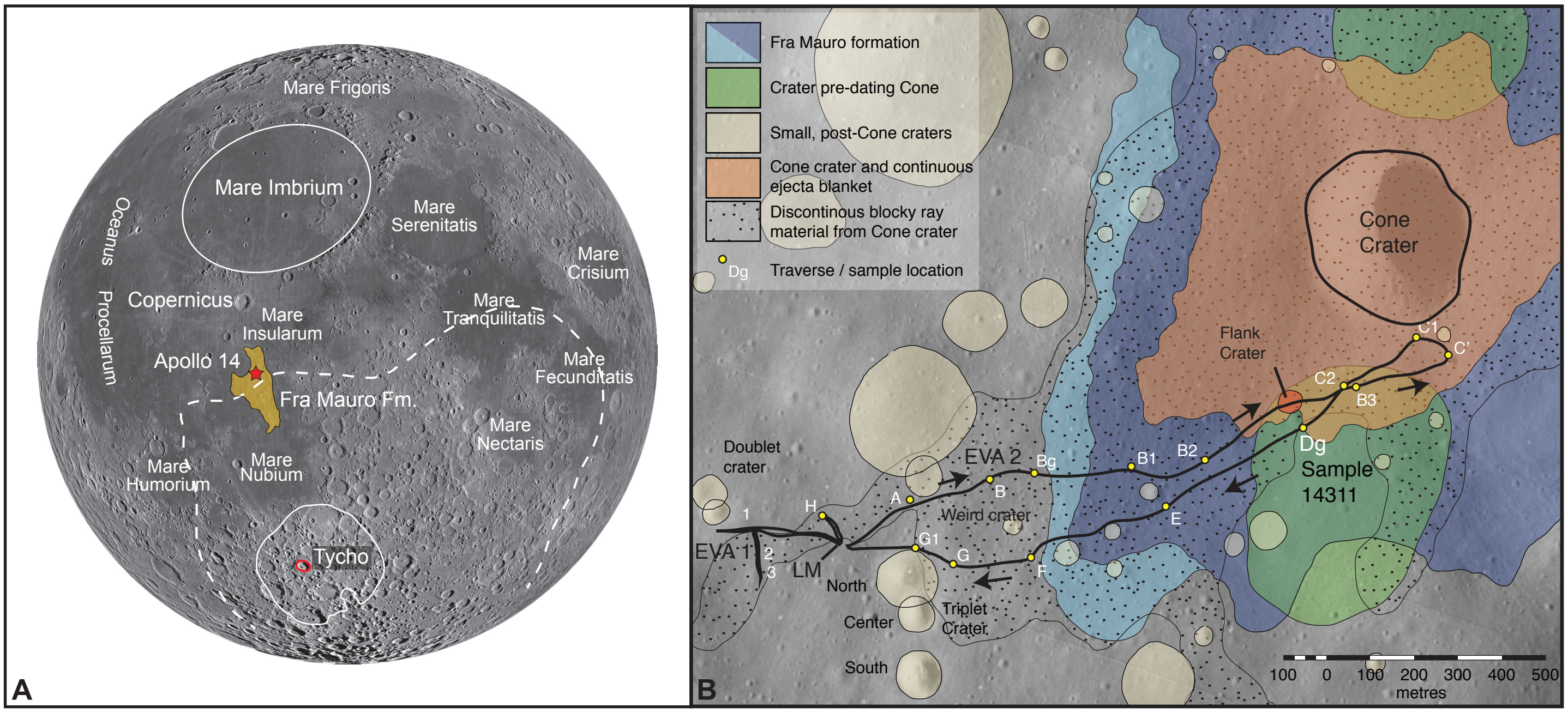

*Figure 1: (2 column width)*

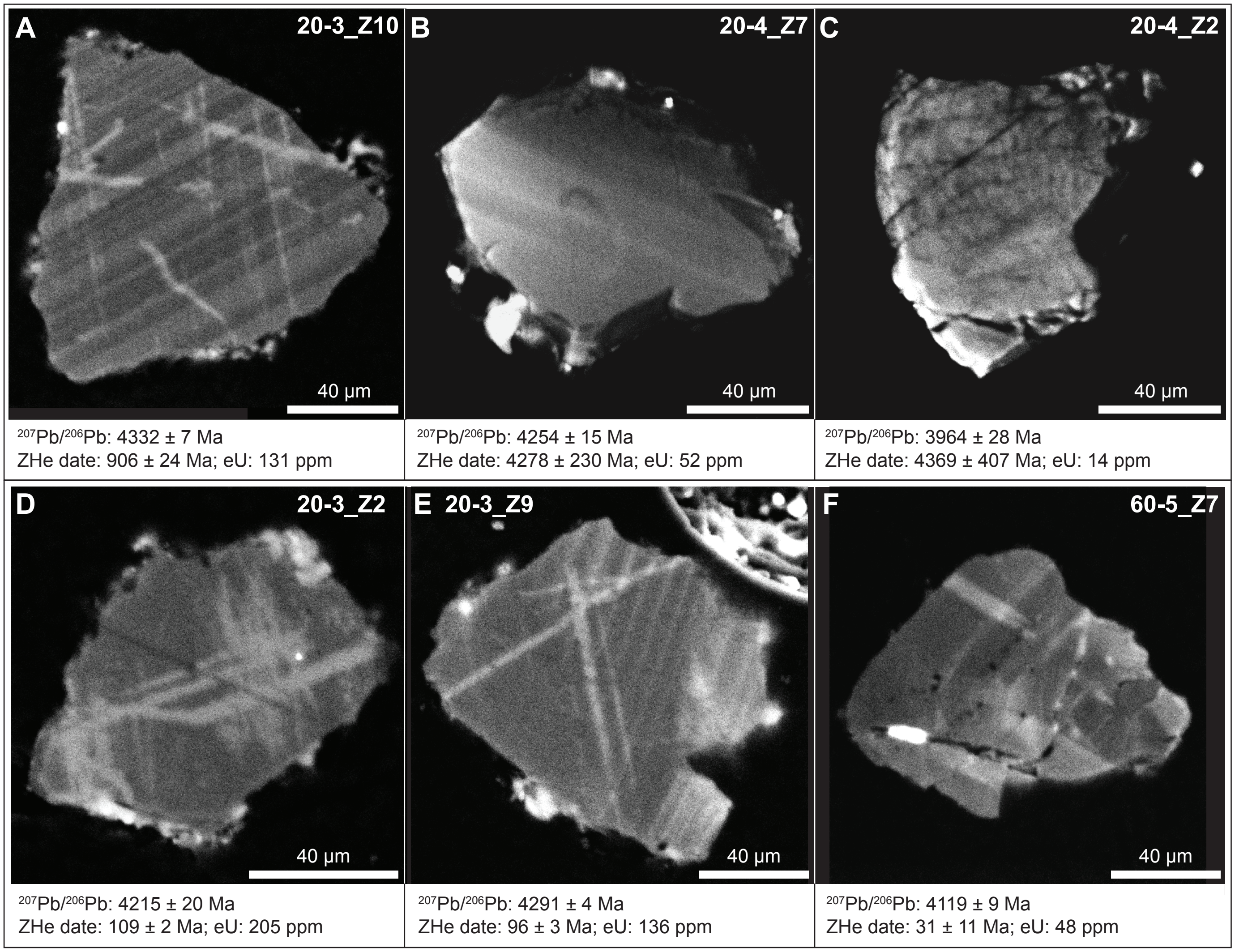

***Figure 2: (double column width)***

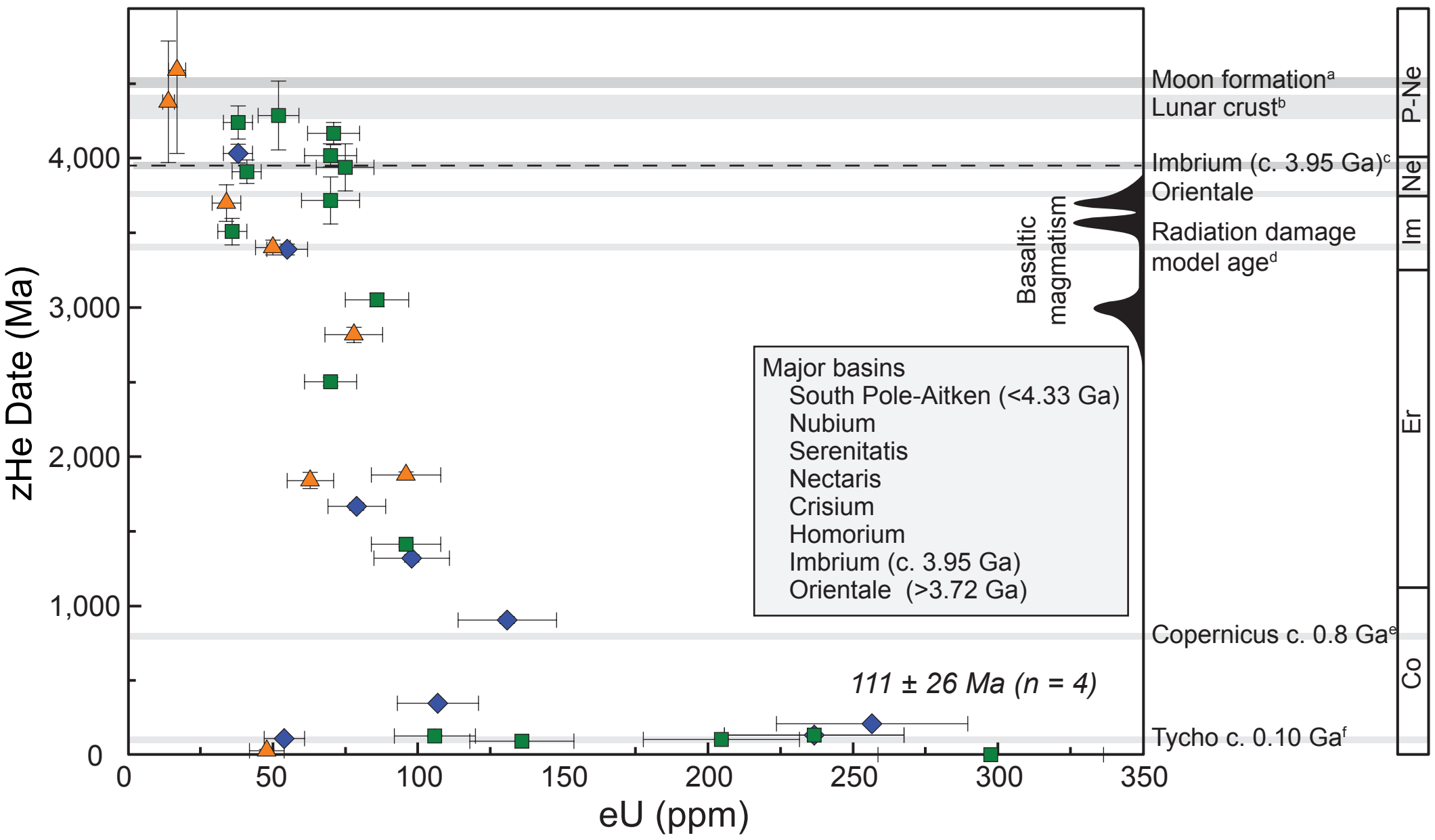

***Figure 3. (1.5 column width)***

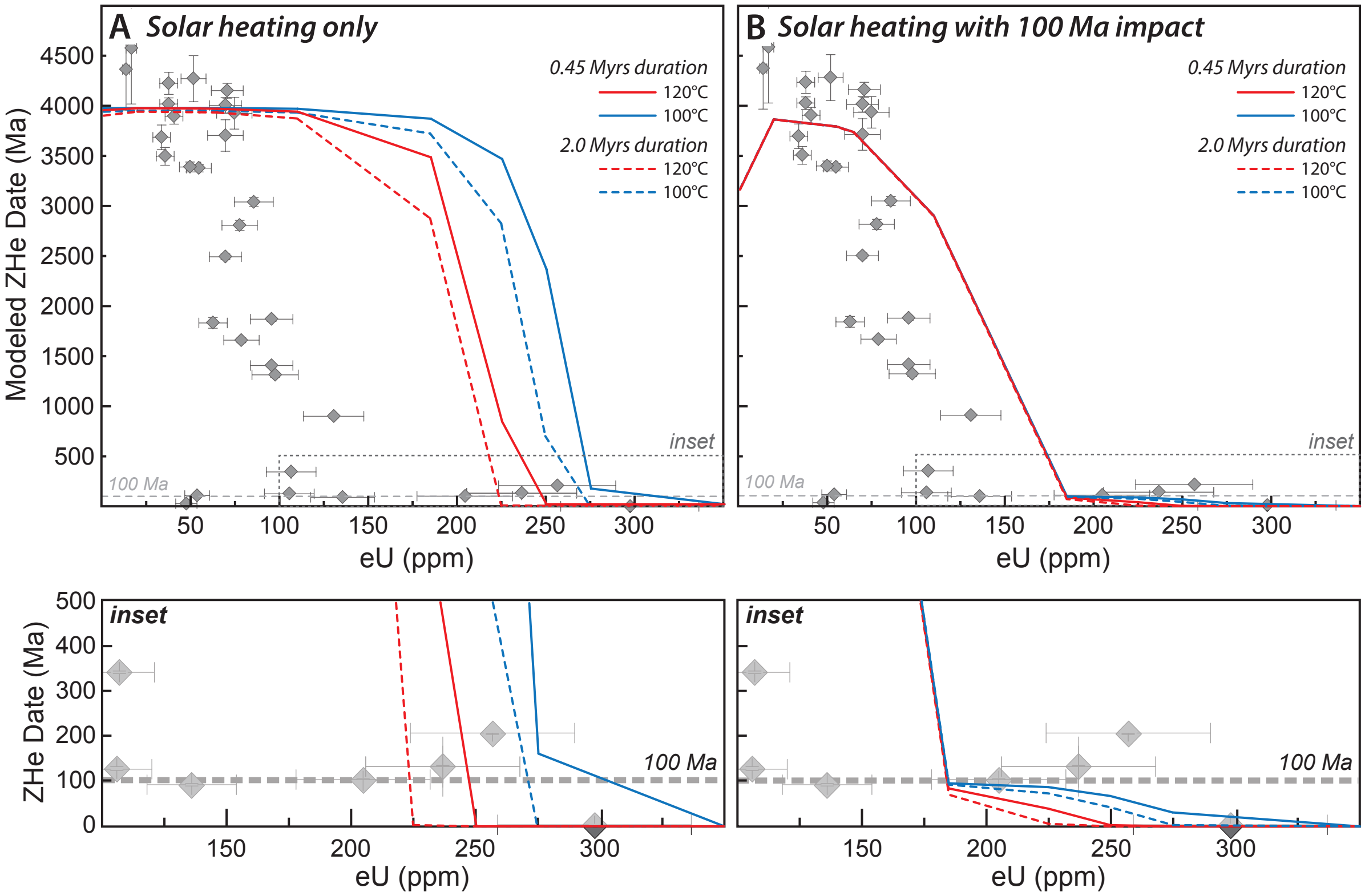

*Figure 4 (2 column width)*

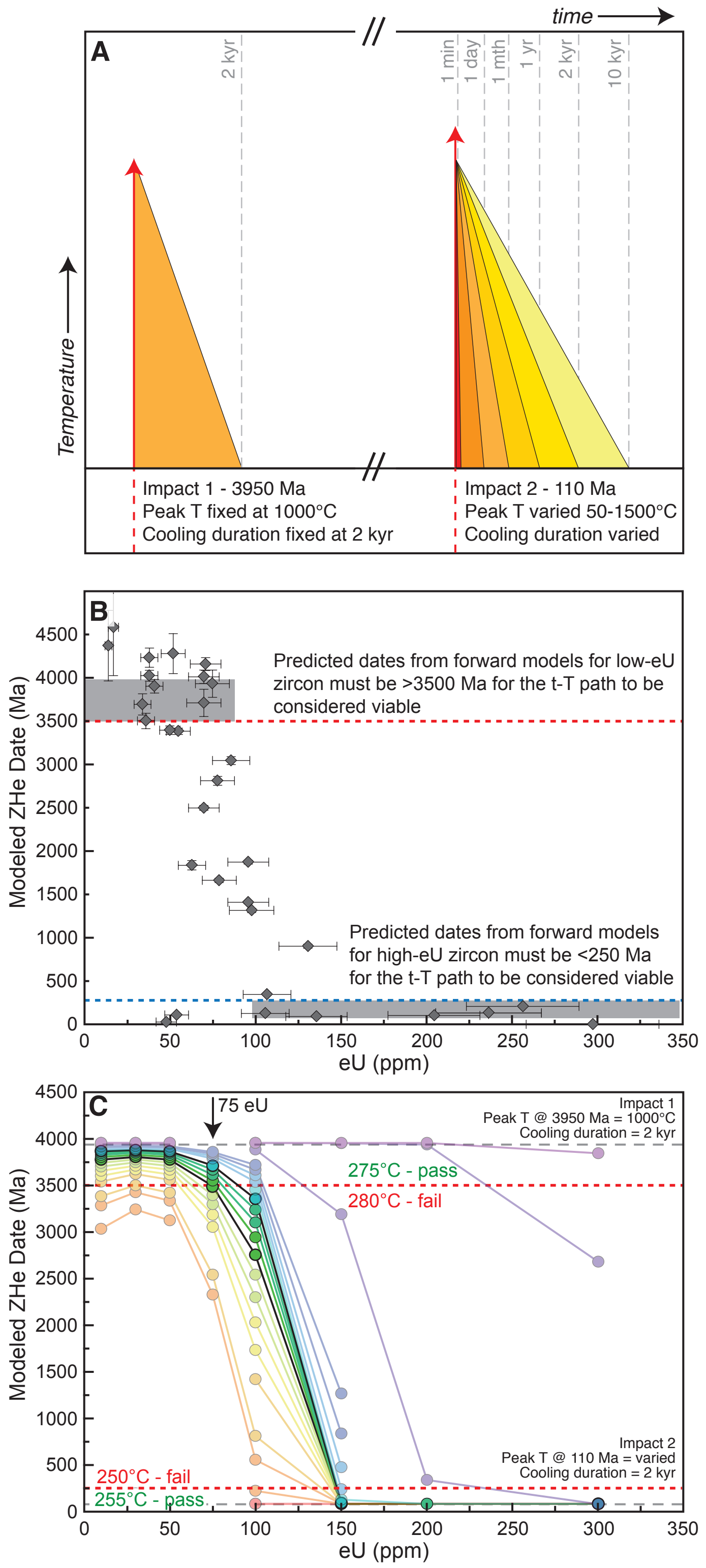

***Figure 5 (1 column width)***

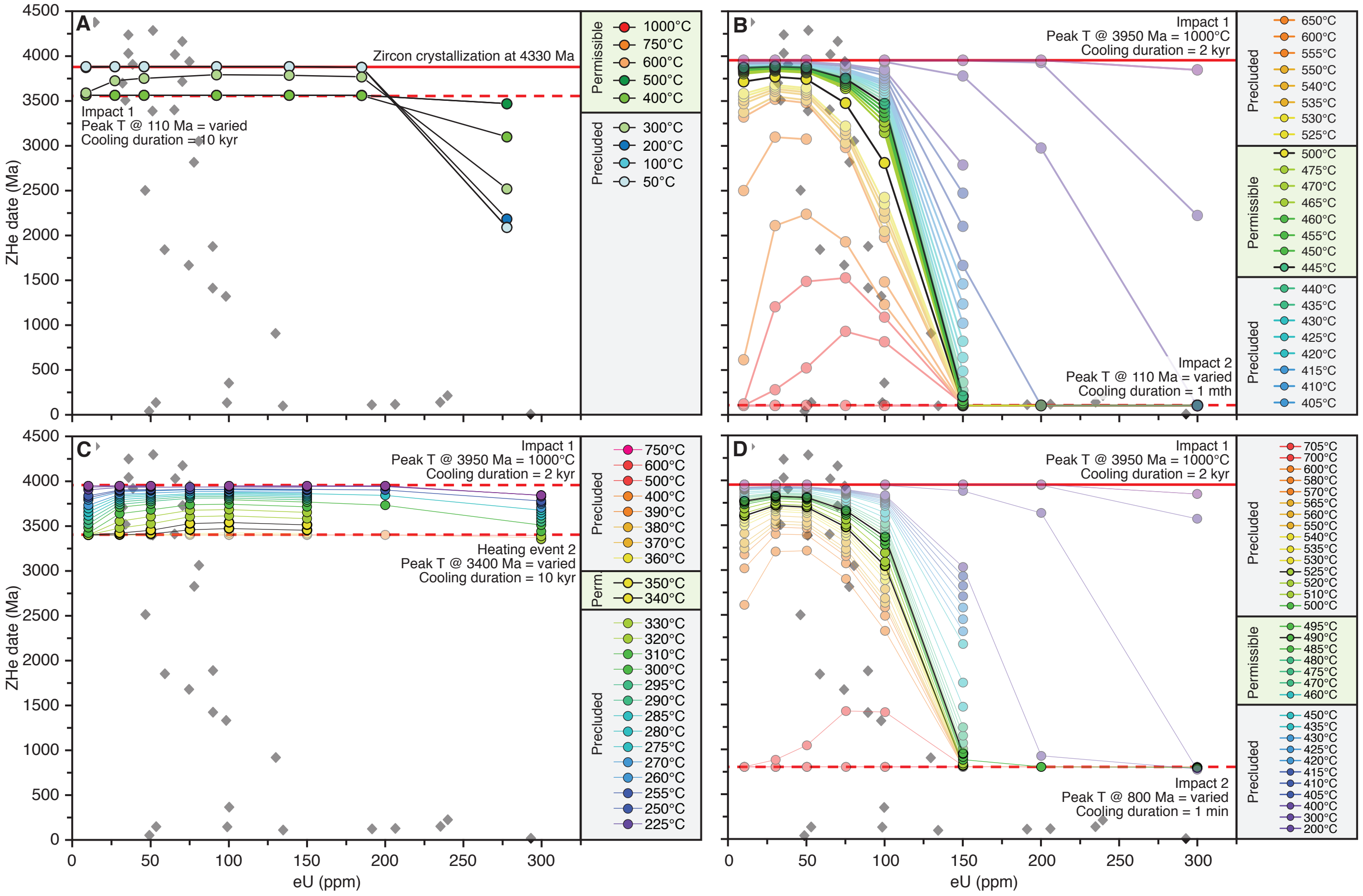

*Figure 6.*

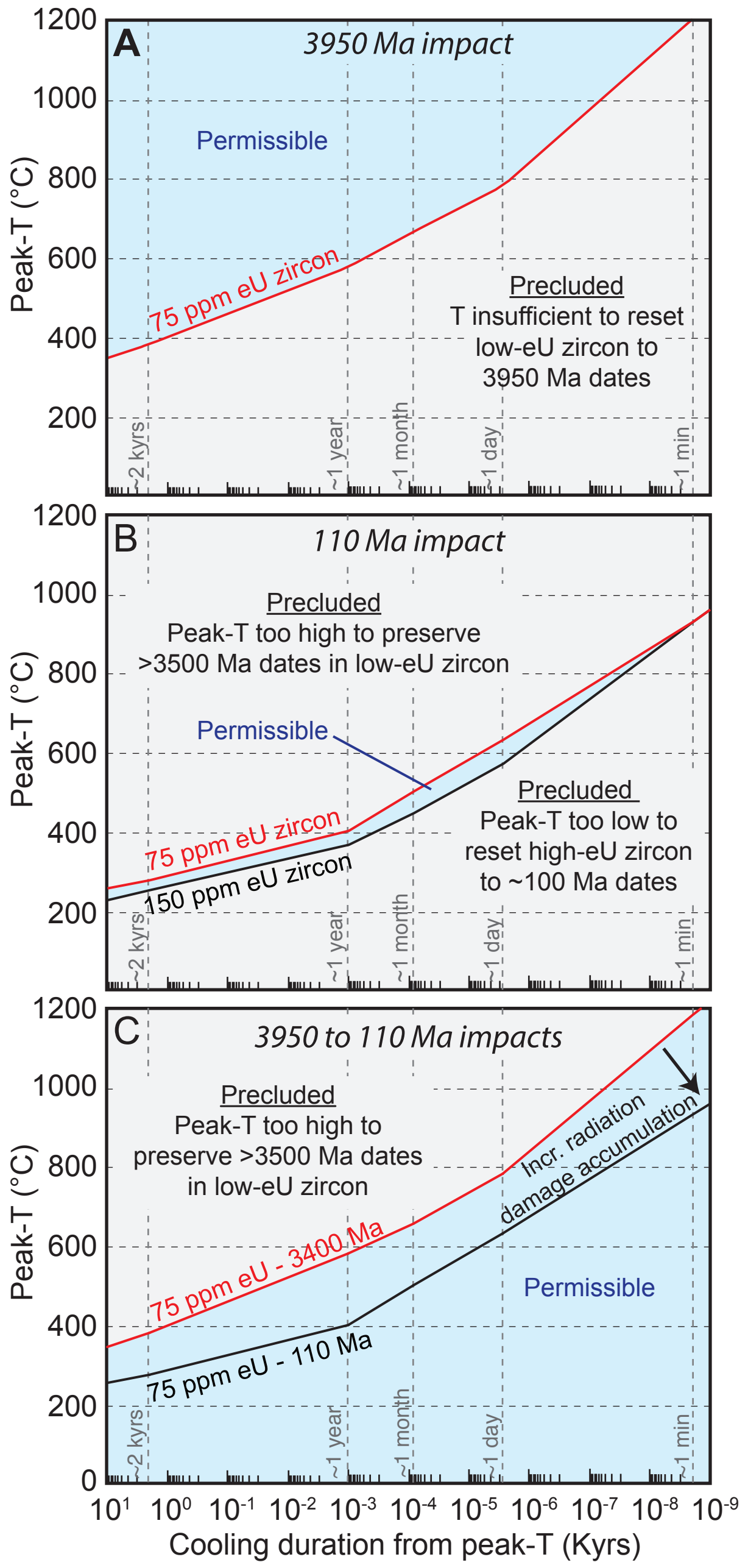

***Figure 7. (1 column width)***

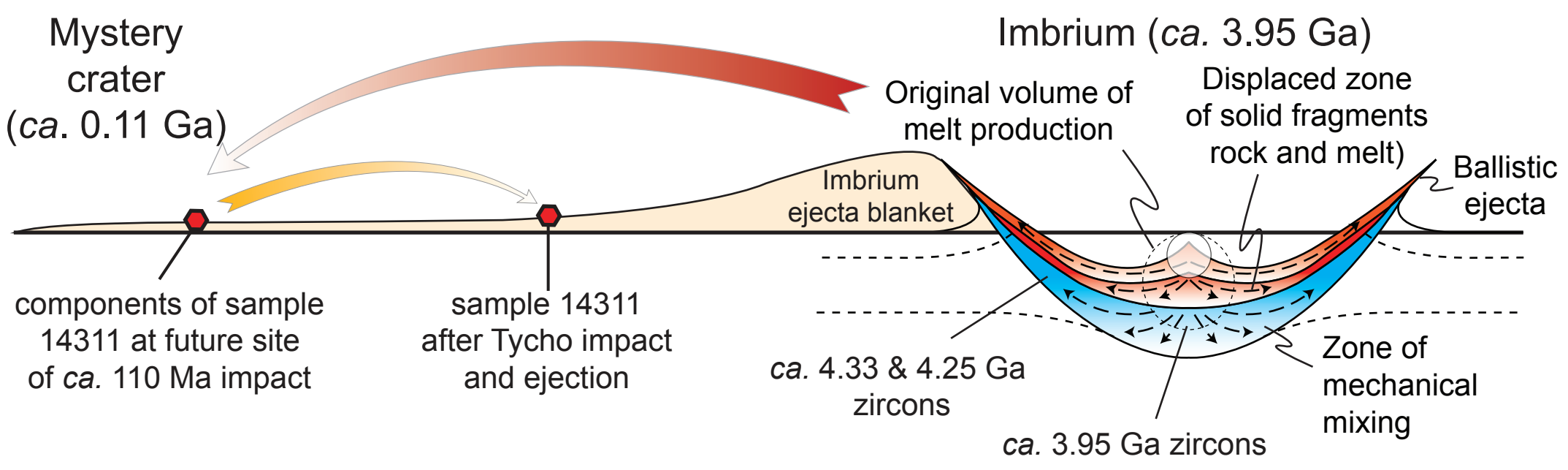

***Figure 8. (1.5 column width)***

*Table 1. Zircon (U-Th)/He data for lunar impact-melt breccia sample 14311.*

| Sample & Grain | Vol.[a] ($mm^3$ x10$^{-3}$) | Mass[b] (µg) | Spherical radius from mass (µm) | U (ppm) | Th (ppm) | Sm (ppm) | eU[c] (ppm) | ±[d] (ppm) | $^4$He (nmol/g) | ± | $^4$He Meas (ncc) | GCR[e] accum. time (Ma) | Cosmo-genic $^4$He[f] (ncc) | % $^4He_{cos}$ of meas. $^4$He | GCR Corr. $^4$He (ncc) | ± | GCR-corr Date (Ma) | ± (Ma) |
|---|---|---|---|---|---|---|---|---|---|---|---|---|---|---|---|---|---|---|
| **14311.20** | | | | | | | | | | | | | | | | | | |
| 20-3_z2 | 0.25 | 1.15 | 39 | 190.7 | 59.0 | 0.0 | 205 | 27 | 123.4 | 0.2 | 3.2 | 661 | 0.061 | 1.92 | 3.1 | 0.00 | 108.7 | 2.2 |
| 20-3_z5 | 0.30 | 1.37 | 41 | 90.2 | 34.7 | 4.8 | 98 | 13 | 795.0 | 1.1 | 24.6 | 661 | 0.090 | 0.37 | 24.5 | 0.03 | 1318.5 | 24.1 |
| 20-3_z9 | 0.39 | 1.83 | 45 | 109.3 | 115.3 | 4.0 | 136 | 18 | 73.4 | 0.1 | 3.0 | 661 | 0.120 | 3.96 | 2.9 | 0.00 | 95.6 | 3.3 |
| 20-3_z10 | 1.41 | 6.55 | 69 | 114.3 | 71.1 | 4.2 | 131 | 17 | 695.2 | 0.5 | 102.5 | 661 | 0.429 | 0.42 | 102.1 | 0.07 | 906.3 | 23.7 |
| 20-3_z13 | 1.01 | 4.72 | 62 | 82.5 | 56.9 | 1.5 | 96 | 12 | 833.1 | 0.9 | 88.4 | 661 | 0.309 | 0.35 | 88.1 | 0.09 | 1412.1 | 16.0 |
| 20-3_z14 | 1.19 | 5.52 | 66 | 91.5 | 64.1 | 1.8 | 107 | 14 | 209.6 | 0.2 | 26.1 | 661 | 0.361 | 1.38 | 25.7 | 0.02 | 350.3 | 3.4 |
| 20-3_z17 | 0.62 | 2.86 | 53 | 29.5 | 28.8 | 0.3 | 36 | 5 | 1042.1 | 0.9 | 67.1 | 661 | 0.188 | 0.28 | 66.9 | 0.06 | 3502.0 | 88.0 |
| 20-4_z1 | 0.17 | 0.78 | 34 | 64.7 | 44.6 | 16.3 | 75 | 10 | 2708.5 | 1.8 | 47.6 | 661 | 0.051 | 0.11 | 47.6 | 0.03 | 3932.5 | 157.3 |
| 20-4_z2 | 0.20 | 0.93 | 36 | 12.0 | 7.6 | 16.7 | 14 | 2 | 622.7 | 0.6 | 13.1 | 661 | 0.061 | 0.47 | 13.0 | 0.01 | 4369.4 | 406.5 |
| $20\text{-}4_z7_h$ | | 1.43 | 42 | 46.0 | 24.4 | 8.0 | 52 | 7 | 2245.0 | 1.0 | 72.0 | 661 | 0.094 | 0.13 | 71.9 | 0.03 | 4278.3 | 229.8 |
| 20-4_z8[h] | | 2.42 | 50 | 240.3 | 245.6 | 6.5 | 298 | 39 | 11.9 | 0.0 | 0.6 | 661 | 0.159 | 24.49 | 0.5 | 0.00 | 5.6 | 1.8 |
| 20-5_z3 | 0.78 | 3.61 | 57 | 67.3 | 51.7 | 0.0 | 79 | 10 | 836.0 | 0.7 | 67.8 | 661 | 0.236 | 0.35 | 67.6 | 0.06 | 1664.8 | 24.0 |
| 20-5_z5 | 0.42 | 1.96 | 46 | 47.7 | 29.5 | 7.6 | 55 | 7 | 1510.2 | 1.4 | 66.5 | 661 | 0.128 | 0.19 | 66.4 | 0.06 | 3382.9 | 35.9 |
| **14311.50** | | | | | | | | | | | | | | | | | | |
| 50-3_z4 | 0.92 | 4.29 | 60 | 35.6 | 21.0 | 4.7 | 41 | 5 | 1448.1 | 4.6 | 139.4 | 661 | 0.281 | 0.20 | 139.1 | 0.44 | 3901.9 | 78.1 |
| 50-3_z6 | 0.44 | 2.05 | 47 | 60.4 | 39.2 | 9.8 | 70 | 9 | 2617.2 | 4.5 | 120.5 | 661 | 0.134 | 0.11 | 120.4 | 0.21 | 4011.3 | 77.5 |
| 50-3_z7 | 0.39 | 1.81 | 45 | 91.6 | 60.3 | 4.0 | 106 | 14 | 78.6 | 0.1 | 3.2 | 661 | 0.119 | 3.69 | 3.1 | 0.00 | 132.2 | 3.9 |
| 50-3_z10 | 1.01 | 4.70 | 62 | 211.6 | 107.4 | 3.8 | 237 | 31 | 180.4 | 0.1 | 19.1 | 661 | 0.308 | 1.61 | 18.8 | 0.01 | 137.7 | 1.8 |
| 50-3_z11 | 1.11 | 5.17 | 64 | 60.2 | 45.8 | 4.9 | 71 | 9 | 2844.9 | 2.7 | 329.7 | 661 | 0.338 | 0.10 | 329.4 | 0.32 | 4157.5 | 74.1 |
| 50-3_z12 | 0.59 | 2.73 | 52 | 29.4 | 18.1 | 4.1 | 34 | 5 | 1083.1 | 1.3 | 66.5 | 661 | 0.179 | 0.27 | 66.3 | 0.08 | 3691.5 | 121.9 |
| **14311.60** | | | | | | | | | | | | | | | | | | |
| 60-2_z4 | 0.42 | 1.96 | 46 | 75.6 | 45.6 | 0.0 | 86 | 11 | 2019.9 | 0.5 | 88.9 | 661 | 0.128 | 0.14 | 88.7 | 0.02 | 3044.6 | 44.6 |
| 60-4_z6 | 0.60 | 2.79 | 63 | 70.5 | 31.8 | 9.1 | 78 | 10 | 1633.8 | 1.0 | 102.5 | 661 | 0.183 | 0.18 | 102.3 | 0.06 | 2811.8 | 50.1 |
| 60-4_z7 | 0.22 | 1.04 | 52 | 62.4 | 32.5 | 53.7 | 70 | 10 | 2296.9 | 1.4 | 53.7 | 661 | 0.068 | 0.13 | 53.6 | 0.03 | 3709.8 | 158.1 |
| 60-4_z9 | 1.10 | 5.12 | 38 | 64.5 | 21.5 | 0.0 | 70 | 9 | 1239.0 | 0.7 | 142.4 | 661 | 0.335 | 0.24 | 142.1 | 0.08 | 2499.0 | 27.6 |
| 60-4_z10 | 1.05 | 4.87 | 64 | 44.7 | 20.7 | 0.0 | 50 | 6 | 1391.8 | 1.2 | 152.1 | 661 | 0.319 | 0.21 | 151.8 | 0.13 | 3395.3 | 48.9 |
| 60-5_z1 | 0.23 | 1.09 | 38 | 14.8 | 9.1 | 0.0 | 17 | 3 | 851.2 | 1.2 | 20.8 | 661 | 0.071 | 0.34 | 20.8 | 0.03 | 4580.5 | 555.4 |
| 60-5_z2 | 0.77 | 3.58 | 57 | 87.9 | 34.2 | 5.5 | 96 | 12 | 1175.7 | 5.1 | 94.5 | 661 | 0.234 | 0.25 | 94.3 | 0.41 | 1874.8 | 22.2 |
| 60-5_z3 | 1.11 | 5.17 | 64 | 32.6 | 22.4 | 1.0 | 38 | 5 | 1424.7 | 6.8 | 165.5 | 661 | 0.339 | 0.20 | 165.1 | 0.79 | 4023.1 | 61.6 |
| 60-5_z4 | 0.20 | 0.94 | 36 | 231.3 | 107.8 | 10.9 | 257 | 33 | 301.5 | 0.4 | 6.4 | 661 | 0.062 | 0.97 | 6.3 | 0.01 | 211.7 | 2.1 |
| 60-5_z5 | 0.22 | 1.04 | 38 | 56.5 | 26.8 | 27.9 | 63 | 8 | 752.2 | 0.9 | 17.5 | 661 | 0.068 | 0.39 | 17.5 | 0.02 | 1838.5 | 54.4 |
| 60-5_z7 | 0.24 | 1.11 | 38 | 43.6 | 18.6 | 0.0 | 48 | 6 | 10.9 | 0.1 | 0.3 | 661 | 0.072 | 26.31 | 0.2 | 0.00 | 31.5 | 10.6 |
| 60-5_z9 | 0.36 | 1.68 | 44 | 46.7 | 32.5 | 1.9 | 54 | 7 | 34.7 | 10.4 | 1.3 | 661 | 0.027 | 2.07 | 1.3 | 0.39 | 114.6 | 2.4 |
| 60-5_z11 | 0.51 | 2.38 | 50 | 33.0 | 21.4 | 0.0 | 38 | 5 | 1592.3 | 14.7 | 85.1 | 661 | 0.156 | 0.18 | 84.9 | 0.79 | 4231.5 | 110.9 |

[a] Volume measured by HR-XCT

[b] Mass calculated using density of pristine zircon (4.65 g/$cm^3$)

[c] eU - effective uranium concentration, combines U and Th weighted by their alpha productivity, computed as [U] + 0.235 * [Th]

[d] uncertainty includes analytical error and uncertainty involving estimated density of pristine versus metamict zircon

[e] GCR - Galactic Cosmic Rays; $^4$He may be produced due to iteractions with GCRs and is referred to here as cosmogenic $^4$He ($^4He_{cos}$)

[f] Cosmogenic $^4$He calculated using maximum production rates for $^4$He from (31), based on estimated composition of the breccia matrix

[g] Age difference between dates calculated using the maximum exposure age (c. 661 Ma, 19) and ZHe date for those grains where ZHe < 661 Ma.

[h] Volume estimate based on grain dimensions

***Table 2. Summary of Temperature-time paths used in HEFTy forward models.***

| | Event Date* | | | | | | | | | | Cooling duration from peak-T to 0°C | | | | | Results in figures† |
|---|---|---|---|---|---|---|---|---|---|---|---|---|---|---|---|---|
| | **4330 Ma** | **4250 Ma** | **3950 Ma** | **3400 Ma** | **3000 Ma** | **2500 Ma** | **2000 Ma** | **1500 Ma** | **800 Ma** | **100 Ma** | **Event 1** | **Event 2** | **Event 3** | **Event 4** | **Temp. range** | |
| ***Impact event forward models*** | | | | | | | | | | | | | | | | |
| Model 1 | | | 1000°C | | | | | | | | 2 kyr | | | | | |
| Model 2 | 1000°C | | X°C | | | | | | | | 33 Myr | 1 Myr - 1 min | | | 1000 - 50°C | Fig. 6a; Fig. S2 |
| Model 3 | 1000°C | | | | | | | | | X°C | 33 Myr | 1 Myr - 1 yr | | | 1000 - 50°C | |
| Model 4 | | 1000°C | | | | | | | X°C | | 2 kyr | 100 kyr - 2 kyr | | | 1000 - 50°C | |
| Model 5 | | 1000°C | | | | | | | | X°C | 2 kyr | 1 Myr - 1 yr | | | 1000 - 50°C | |
| Model 6 | | | 1000°C | | | | | | | X°C | 2 kyr | 1 Myr - 1 min | | | 1000 - 50°C | Fig. 5c, 6b,c; Fig. S3 |
| Model 7 | | | 1000°C | X°C | | | | | | | 2 kyr | 10 kyr - 1 min | | | 1500 - 50°C | Fig 5d; Fig. S4, S5 |
| Model 8 | | | 1000°C | | X°C | | | | | | 2 kyr | 10 kyr - 1 min | | | 1500 - 50°C | |
| Model 9 | | | 1000°C | | | X°C | | | | | 2 kyr | 10 kyr - 1 min | | | 1500 - 50°C | |
| Model 10 | | | 1000°C | | | | X°C | | | | 2 kyr | 10 kyr - 1 min | | | 1000 - 50°C | |
| Model 11 | | | 1000°C | | | | | X°C | | | 2 kyr | 10 kyr - 1 min | | | 2000 - 50°C | |
| Model 12 | | | 1000°C | | | | | | X°C | | 2 kyr | 100 kyr - 1 min | | | 1000 - 50°C | Fig 5e; Fig. S6 |
| Model 13 | | | 1000°C | 300°C | | | | | | X°C | 2 kyr | 2 kyr | 10 kyr - 1 min | | 1000 - 50°C | |
| Model 14 | | | 1000°C | 500°C | | | | | | X°C | 2 kyr | 1 yr | 10 kyr - 1 min | | 1000 - 50°C | |
| Model 15 | 1000°C | | 300°C | | | | | | X°C | | 33 Myr | 2 kyr | 1 Myr - 1 yr | | 1000 - 50°C | |
| Model 16 | 1000°C | | 1000°C | | | | | | | X°C | 33 Myr | 10 kyr | 1 Myr - 1 min | | 1000 - 50°C | |
| Model 17 | 1000°C | | 300°C | | | | | | | X°C | 33 Myr | 2 kyr | 1 Myr - 1 min | | 1000 - 50°C | |
| Model 18 | | 1000°C | 300°C | | | | | | 300°C | X°C | 2 kyr | 2 kyr | 100 kyr | 100 kyr - 1 yr | 1000 - 50°C | |
| Model 19 | | 1000°C | 300°C | | | | | | 300°C | X°C | 2 kyr | 2 kyr | 2 kyr | 100 kyr - 1min | 1000 - 50°C | |
| Model 20 | | 1000°C | 300°C | | | | | | 300°C | X°C | 2 kyr | 2 kyr | 1 yr | 100 kyr - 1min | 1000 - 50°C | |
| Model 21 | | 1000°C | 300°C | | | | | | 300°C | X°C | 2 kyr | 2 kyr | 10 kyr | 100 kyr - 1 yr | 1000 - 50°C | |
| Model 22 | | 1000°C | 300°C | | | | | | 300°C | X°C | 2 kyr | 10 kyr | 1 yr | 100 kyr - 1 yr | 1000 - 50°C | |
| ***Solar cycling models*** | | | **3950 Ma** | | | | | | | | | | | | | |
| Solar Cyling @ 0.45 - 25Ma | | | 1000°C | | | | | | | | 2 Kyrs | 5 min | | | 107°C | Fig. 4 |
| Solar Cyling @ 0.45 - 25Ma | | | | | | | | | 400°C | | | 1 yr | | | 107°C | |
| Solar Cyling @ 0.45 - 25Ma | | | | | | | | | | 400°C | | 1 yr | | | 107°C | |

*All heating times to peak-T are ≤5 minutes. Most model histories used maximum peak temperatures of 1000°C – higher temperatures result in complete resetting of zircons at all eU values during events as short as a few minutes and so were not investigated. Zircon grainsize used in models: 100 μm radius (assuming measure grain sizes are smaller than original, in situ grains due to fragmentation during mechanical separation); Zircon eU compositions: 10, 30, 50, 75, 100, 150, 200, 300 (some models excluded 200 eU due to time constraints). Where T°C is given under an event date, peak temperature was fixed across models; where X°C is given, peak-T was varied between models. Fig. Sx = Supplementary Information Figure Sx*

# Supplementary Files

## Kelly et al.: Late accretion to the Moon recorded in zircon (U-Th)/He thermochronometry

## Includes:

Analytical methods and data reduction techniques
Supplementary Figures S1-S9
Supplementary Tables S1-S6

**Methods**

**Grain characterization by HRXCT.** In contrast to the euhedral morphology of crystals more commonly used in terrestrial (U-Th)/He thermochronometry, the lunar grains used in this study were predominantly fragments formed during grain separation. Therefore, to determine grain mass – used to calculate U, Th and Sm concentrations from ICP-MS data – all zircon grain shapes and volumes were determined using high-resolution X-ray computed tomography (HRXCT) at the University of Texas at Austin High-Resolution X-ray Computed Tomography Facility (UTCT) and Colorado School of Mines Micro X-Ray CT Facility (CSM-CT). Imaging at UTCT was conducted using the XRADIA MICROXCT scanner using a 4X objective at 100 kV and 10W. Imaging was carried out throughout a 360° rotation with an acquisition time of 2 seconds each for 45 frames. Xradia Reconstructor was used to reconstruct raw x-ray data to 16bit TIFF images, with a voxel size of 4 μm. Imaging at CSM-CT was conducted using the XRadia MicroXCT-400 scanner using a 4X objective at 150 kV and 4.8 W. Scans were obtained for a 188° rotation with an angular increment of 0.5° (377 frames) at 30 s exposure time for each frame. The CT volume was reconstructed to 16bit RAW TIFF files at a 4.35 μm voxel size. Reconstructed X-ray data were processed using the computer program "Blob3D" (v2015, Ketcham et al., 2005b), which enables separation of discrete objects within solid samples for, in the case of the current study, volume estimates of the individual, irregularly shaped zircon grains.

**Sample Preparation.** Following comprehensive characterization by optical microscopy and scanning electron microscopy (CL, BSE), select zircon grains were carefully pried from the epoxy mounts using a needle under a drop of ethanol, and transferred by tweezers or graduated pipette to an ethanol-filled petri dish for further optical imaging using a Leica M165 binocular microscope equipped with a calibrated digital camera and capable of both reflected and transmitted, polarized light. Dimensions of each plucked grain were measured for comparison with HRXCT images and data. In rare cases where fractured zircon grains fragmented during plucking, images and measurements allowed for comparison to in situ measurements (optical images and HRXCT) and appropriate corrections to volume data made where small pieces of zircon were inferred lost. Plucked zircon grains were individually placed into small Nb tubes that are then crimped on both ends. Petri dishes were emptied and cleaned between transfer and packing of each zircon grain.

**(U-Th)/He analysis.** All ZHe analyses were acquired at the University of Colorado at Boulder, using the zircon degassing and dissolution methods described in Stanley & Flowers (2016). To ensure background $^4$He was low, no analysis of lunar grains was conducted in the week following analysis of minerals with high He abundances (e.g., titanite), and low backgrounds were verified by a series of cold blanks run during the overnight pumpdown. Measured $^4$He abundances were well above detection limits and background $^4$He levels (<0.001 ncc). Measured U, Th, and Sm contents (in absolute quantities, mols) were converted to concentrations (in ppm) using calculated masses of each grain (from HRXCT volume data) using the ideal zircon density of 4.65 g/cm$^3$. Compositional variation in zircon (REE, U, Th) will lead to a minor increase in this ideal density. However, the most pronounced deviation from ideal is expected from accumulated radiation damage, which leads to a relative decrease in density due to volume expansion. Swelling from localized defects (e.g., particle tracks) has been estimated to be up to 5%, while swelling due to recoil damage may be up to 13% (Salje et al, 1999), leading to a reduction in density to ~4.42 g/cm$^3$ and ~4.05 g/cm$^3$, respectively. Uncertainties on U-Th-Sm concentrations were calculated to include a maximum possible density variation (13%) along with analytical uncertainty from ICP-MS analysis. Uncertainties on eU values (including that used in figures) represent a linear combination of these two uncertainties.

ZHe dates and all associated data were calculated on a custom, in-house CU TRaIL spreadsheet. Grain shapes of the lunar zircons suggested they were grain fragments produced during initial sample preparation, with few, if any, original grain surfaces preserved. Therefore, no correction for α-ejection was applied to calculated ZHe dates. All ZHe dates and associated element concentration data were calculated using an iterative approach that is required for ages >1 Ga (Ketcham et al., 2011). Individual dates are reported at 1σ uncertainty in the text, figures and tables. Uncertainties on dates only include propagated analytical uncertainties for measurements of He, U, Th and Sm. Uncertainties typically also include those propagated from alpha-ejection correction and for grain volume calculations which are based on manual measurement of zircon grain widths and lengths. However, neither were performed in this study and so not included in uncertainty calculation.

**Estimating cosmogenic $^4$He production.** All measured $^4$He ($^4He_{TOT}$) abundances were corrected for cosmogenic $^4$He ($^4He_{COS}$) produced by galactic cosmic rays, which is a function of the cosmogenic isotope production rate, the mass of the target (e.g., mineral or rock) and the accumulation time (typically estimated from exposure ages). Contribution from solar cosmic rays and He particles in solar wind is considered negligible (Crozaz et al., 1972; Rao et al., 1993), so no correction was applied.

The contribution of $^4He_{COS}$ to $^4He_{TOT}$ was calculated using the experimentally derived $^4He_{COS}$ production rate ($P_4$) values from (Leya et al., 2004) for production in a gabbroic matrix (maximum $P_4 \approx 8.71 \times 10^{-8}$ cm$^3$ STP/g Ma), and through extrapolation (Farley et al., 2006) from experimental values, for production in zircon (Si: maximum $P_4 \approx 8.63 \times 10^{-8}$ cm$^3$ STP/g Ma; Zr: maximum $P_4 \approx 1.96 \times 10^{-8}$ cm$^3$ STP/g Ma; to achieve a maximum zircon $P_4 \approx 6.67 \times 10^{-8}$ cm$^3$ STP/g Ma, for a lunar proton flux of $J = 4.54$ cm$^{-2}$ s$^{-1}$). Stopping distances of spalled $^4$He are poorly known (stopping distance of spalled $^3$He is ~50 µm, Farley et al., 2006), and it is highly probable that some or all $^4He_{COS}$ present in the zircon grains was implanted from the matrix. Therefore, correction of $^4He_{TOT}$ for $^4He_{COS}$ contribution was made using the experimental production values for the gabbroic matrix composition corrected for the composition of lunar breccia sample 14311,67 (Eugster, 1988; Scoon, 1972), achieving a maximum $P_4$ value of $9.91 \times 10^{-8}$ cm$^3$ STP/g Ma (see summary in Table S2).

Accumulation times used in $^4He_{TOT}$ correction were based on the published exposure age of sample 14311 (*ca.* 661 Ma, Crozaz et al., 1972). However, assuming that diffusivity of radiogenic $^4$He ($^4He_{RAD}$) and $^4He_{COS}$ is identical, the proportion of $^4He_{COS}$ lost during any heating event subsequent to *ca.* 661 Ma will be proportional to total He loss. Therefore, $^4He_{COS}$ calculated for any zircon with a ZHe date younger than *ca.* 661 Ma will be overestimated. Calculated $^4He_{COS}$ values using the exposure age are between 0.027 and 0.429 ncc. For all but 2 grains this represents between 0.1 and 4% of $He_{TOT}$, with two outlier grains (20-4_Z8 and 60-5_z7) having higher values (24.5% and 26.3%, respectively). When $^4He_{COS}$ is recalculated for grains with ZHe dates younger than *ca.* 661 Ma (9 grains; Table S1, S2), all calculated $^4He_{COS}$ values are less than 1.7% of $^4He_{TOT}$. All plots of ZHe dates presented here are calculated from He values corrected using the maximum value for $^4He_{COS}$. This leads to a maximum overcorrection of 0.6-10 Myrs (most <3 Myrs) for 8 of the 9 grains with ZHe younger than *ca.* 661 Ma, with one grain under corrected by 1 Myr.

**Forward Modeling - method.** The computer program HeFTy (Ketcham et al., 2005a) is most commonly used to calculate inverse models to test possible time-Temperature histories from known age and thermal constraints. However, the manual interface for inverse models precludes input of short-duration (thousands of years and younger) events that correspond to impacts, within histories spanning up to 4 Gyrs. Therefore, forward modeling using HeFTy, which allows digital input of tT parameters, was conducted. Forward models were performed, using the zircon radiation damage model of (Guenthner et al., 2013), which describes evolving diffusivity with time during radiation damage accumulation and annealing. Diffusion experiments have demonstrated diffusion rates in zircon are directionally asymmetric (diffusion parallel to the c-axis of zircon is faster than diffusion orthogonal to the c-axis (Farley, 2007; Cherniak e al., 2009; Guenthner et al., 2013), such that diffusivity will initially decrease as radiation damage blocks fast diffusion pathways parallel to the c-axis of zircon (see discussion in Guenthner et al., 2013; Ketcham et al., 2013). With continued damage accumulation, diffusion behavior reaches a point where diffusivity rapidly increases, likely related to crossing the first percolation threshold (FPT) and development of interconnected networks of damaged zones (Holland, 1954; Hurley, 1954; Reiners, 2005; Ketcham et al., 2013). The damage model by (Guenthner et al., 2013) used in the forward models also uses the fission-track annealing model of (Yamada et al., 2007) to account for partial to complete annealing of accumulated radiation damage during thermal events.

**Forward Modeling – uncertainty in the behavior of highly radiation damaged zircon grains.** The kinetics of zircon He diffusion and damage annealing are best constrained for zircon with low to moderate damage levels (e.g., Reiners et al., 2002, 2004; Wolfe and Stockli, 2010; Guenthner et al., 2013). Fewer data are available for highly radiation damaged zircon with damage levels above the FPT (Zhang et al., 2000: $2.2 \times 10^{18}$ $\alpha$/gram, Salje et al., 1999: $3.5 \times 10^{18}$ $\alpha$/gram). Currently available kinetic models for damage accumulation and annealing (e.g., Guenthner et al., 2013) rely on the annealing behavior of fission tracks (e.g., Yamada et al., 2007), which is interpreted to initiate at temperatures >250°C. Although recent studies suggest that damage from both alpha- and fission-decay affect $^4$He diffusivity (Ketcham et al., 2013), fission-tracks are the only form of radiation damage in zircon for which annealing kinetics are documented on geologic timescales (e.g., Guenthner et al., 2013). Fission-track and alpha-damage annealing in apatite are correlated, and both scale with the observed radiation damage induced changes in $^4$He diffusivity (Shuster and Farley, 2009). Fission-track annealing is therefore used as a proxy for total radiation damage annealing in the current formulation of the ZRDAAM (Guenthner et al., 2013) because it can reasonably approximate damage levels over geologic timescales.

For the lunar zircon grains analyzed in this study (eU = 10-296 ppm), all grains with eU ≥ 100 ppm will be at or above the published minimum estimates for the FPT after 3950 Myrs of damage accumulation (assumed maximum due to full annealing during the Imbrium event), with a maximum damage estimate of ~$6.7 \times 10^{18}$ $\alpha$/gram (see Table S4). This range of zircon compositions corresponds to the grains that are dominated by the *ca.* 110 Ma ZHe dates, and the uncertainty in He retentivity at high damage levels bears on interpreting the significance of these results. The modeled minimum temperature constraints on the 110 Ma event are specifically based on the resetting behavior of >150 ppm eU zircon. If we assume models *overestimate* He retentivity, our minimum-temperatures required to see this signal in the high-eU zircons will also be overestimated. In contrast, if retentivity is *underestimated*, then we also underestimate the temperatures required to generate this signal. Importantly, the *upper*-temperature constraints on the *ca.* 110 Ma event rely on behavior of *low-eU* grains (specifically eU <100 ppm), all of which were below the FPT at *ca.* 110 Ma. All calculated alpha-doses for the low-eU zircons are therefore within the damage range that is well constrained within the ZRDAAM, and interpretations based on this array of zircon compositions are robust within the stated limits of the models.

An additional caveat is uncertainty in our understanding of annealing behavior, including the temperature range and timescales over which it occurs, and if there is differential annealing behavior or mechanisms across different damage states. HeFTy models, which output alpha-dose estimates that include radiation damage recovery, suggest that partial annealing occurred for the predicted maximum temperatures for a *ca.* 110 Ma event (Table S5). If the ZRDAAM underestimates annealing, this would increase retentivity in high-eU grains, and so increase the minimum temperatures required during a *ca.* 110 Ma event, and offset underestimates from poorer retentivity discussed above. Regardless, ongoing advances in understanding high-damage zircon He diffusion kinetic and annealing behavior can be used to refine our interpretations of this lunar dataset in the future.

**Forward Modeling – solar cycling.** To enable forward modeling of solar cycling effects on ZHe dates, the "effective diffusion temperature" (EDT) approach was employed (see Shuster and Cassata, 2015; Tremblay et al., 2014). Mean diffusivities and EDT values were calculated using published diffusivities and activation energies for zircon grains with a range of accumulated radiation damage (see Guenthner et al., 2013) and for temperatures from 120°C (maximum lunar surface temperature) down to 80°C, at 10°C increments (Table S3). Because alpha-dose reflects both U+Th content of a zircon and the duration of damage accumulation, equivalent eU values for each EDT value were calculated by iterating eU with alpha-dose over an accumulation period of 3950 Myrs. Therefore, for input into HeFTy forward models we were able to calculate EDT values for specific eU values and peak temperatures of solar heating.

A gap exists in the published diffusion data for zircon with calculated alpha-doses between 400 and 800 $x10^{16}$ α/g. Therefore, model values for EDT for this part of the damage spectrum were calculated through interpolation of the published diffusivity and activation energy data, and calculated mean diffusivity and EDT values. Iteration of values in forward models demonstrated that the degree of radiation damage after 3950 Myrs for zircon grains in the range eU=200-275 ppm was more strongly controlling of results than chosen EDT, reducing the potential effect of uncertainties in estimates of EDT for zircons in this eU range.

Solar heating forward models were run for heating durations of 0.45 Myrs (minimum estimated *at-surface* exposure period), 1 Myrs, and 2 Myrs (maximum period of *at-surface* exposure period) for sample 14311. Two sets of models were run:

1) a single impact-heating event at 3950 Ma (peak temperature of 1000°C, cooling duration of 2 kyr), followed by residence at zero degrees until initiation of solar heating,

2) an impact-heating event at 3950 Ma (peak temperature of 1000°C, cooling duration of 2 kyr), residence at zero degrees until a second impact event at 110 Ma (peak temperature of 400°C, cooling duration of 1 yr), followed by further residence at zero degrees until initiation of solar heating.

Solar-heating in both cases involved holding grains at the EDT for the required heating duration, before dropping temperature to zero degrees at the end of each model.

**Forward Modeling – impacts.** In order to develop best-case interpretations of the distribution of ZHe dates from 14311, a series of forward models were run to test the influences of potential thermal scenarios on ZHe dates. Model scenarios were initially based on known impact events on the lunar nearside, but extended to evaluate the entire lunar history from 3950 Ma.

Within an impact environment, samples may experience different heating and cooling regimes (i.e., peak-temperature attained, rates of heating and cooling) depending on the size of the impact (high energy, high melt volume generated), location of a sample in the target zone relative to the impactor (within the melting zone or the zone of mechanical mixing, Osinski et al., 2011), and if excavated from the impact crater, relative position within the ejecta blanket (both distance from the impact and depth with in the ejecta blanket). At a first approximation, these factors influence the melt-solid ratio of a sample, which in turn will control the rate of thermal equilibration between clasts and melt, and so the peak temperatures attained and the rate of cooling. It is probable that within a large ejecta blanket like that formed from the Imbrium impact, a population of zircon grains could have experienced temperatures in excess of 2000°C in areas with a high melt-solid ratio (e.g., Grieve et al., 1977; Prevec and Cawthorn, 2002). A sample deep within such an ejection blanket may also cool slowly due to the potential for convection within the blanket (durations of 50-100 kyrs, Prevec and Cawthorn, 2002) and the likelihood of insulation by overlying materials. In contrast, samples lying closer to the surface of an ejecta blanket or ejected further from the impact crater are more likely to have lower melt-solid ratios, and so will attain lower peak temperatures. For example, an addition of 60% of cold clasts will lead to a rapid drop in *melt* temperature from 2000°C to ~1000°C within less than a minute due to thermal equilibration and a reduction in the peak temperatures attained by the clasts themselves (Onorato et al., 1977). A higher clast content and absence of insulation by blanketing may lead to a further decrease in peak temperatures attained, and a concomitant increase in the rate of cooling (years or less, Onorato et al., 1977; Prevec and Cawthorn, 2002). Therefore, model parameters were chosen to test the effect on ZHe dates of: 1) peak temperatures attained, 2) duration of cooling from peak temperature, 3) time between heating/impact events (increasing degree of radiation damage accumulation), and 4) number of impact events.

Simulated impact events in the temperature-time (T-t) histories for all models assumed rapid heating (≤5 minutes) to peak temperatures that varied from 1500°C to 50°C, and linear cooling trajectories over durations of >10 kyrs to 1 minute (time taken to cool from peak-temperature to 0°C). A summary of all T-t models run is provided in Table S3, with output of select models presented in Figs. S3-7. Note, where testing event histories with multiple impacts, peak temperatures and cooling durations of only one impact event was varied, while others remained fixed.

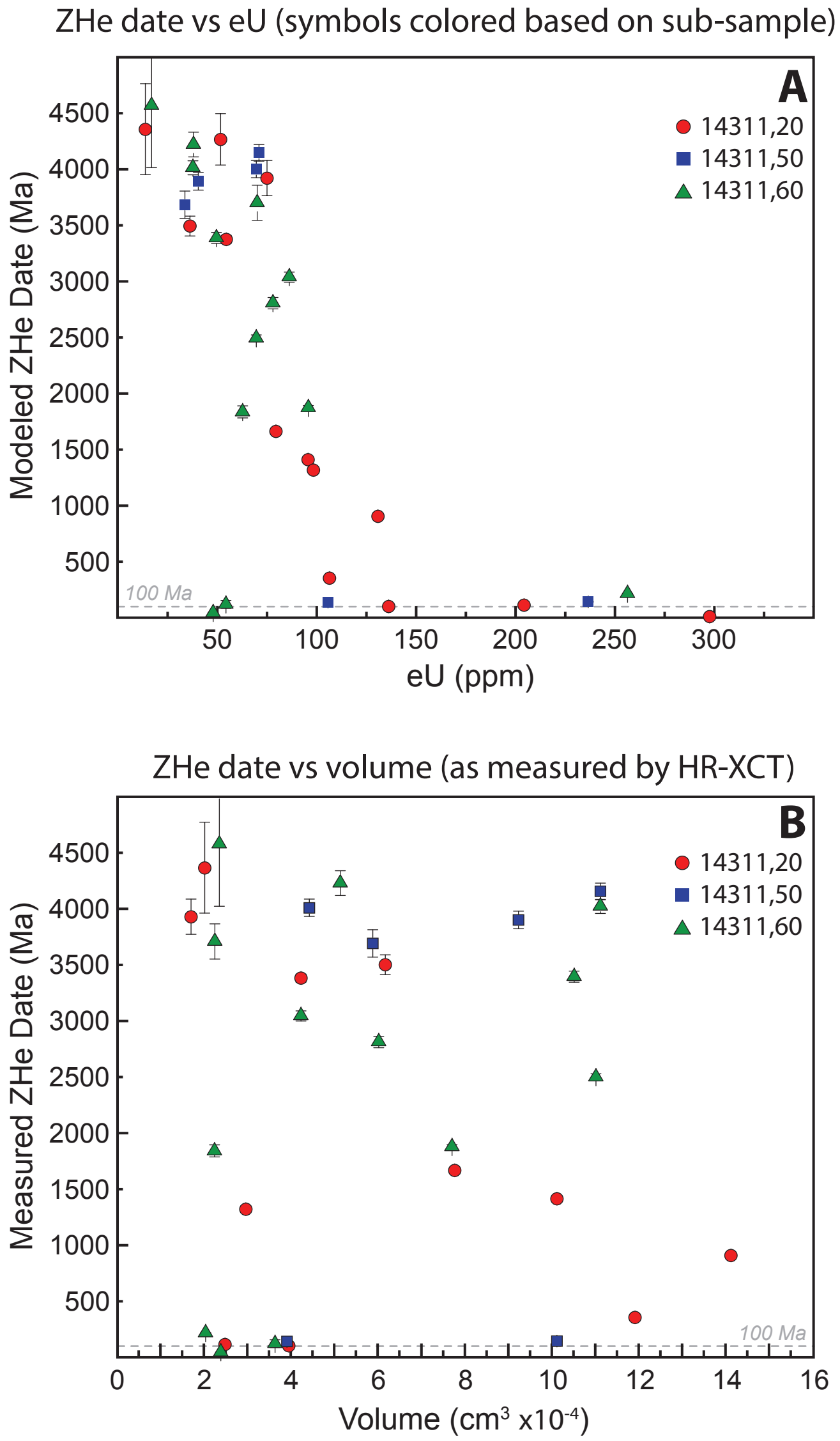

***Supplementary Figure S1.*** Measured ZHe data from sample 14311, symbols and colors coded based on sub-samples 14311,20 (red circles), 14311,50 (blue squares) and 14311,60 (green triangles). a) Date-eU plot illustrating that there is no systematic difference in dates or eU values between subsamples, and therefore likely no bias due to solar heating on a single surface of the sample. b) ZHe date vs grain volume (calculated from HR-XCT data) illustrating no systematic relationship between grain size and ZHe date. We interpret this to suggest that our forward modeling using a single grain-size is valid.

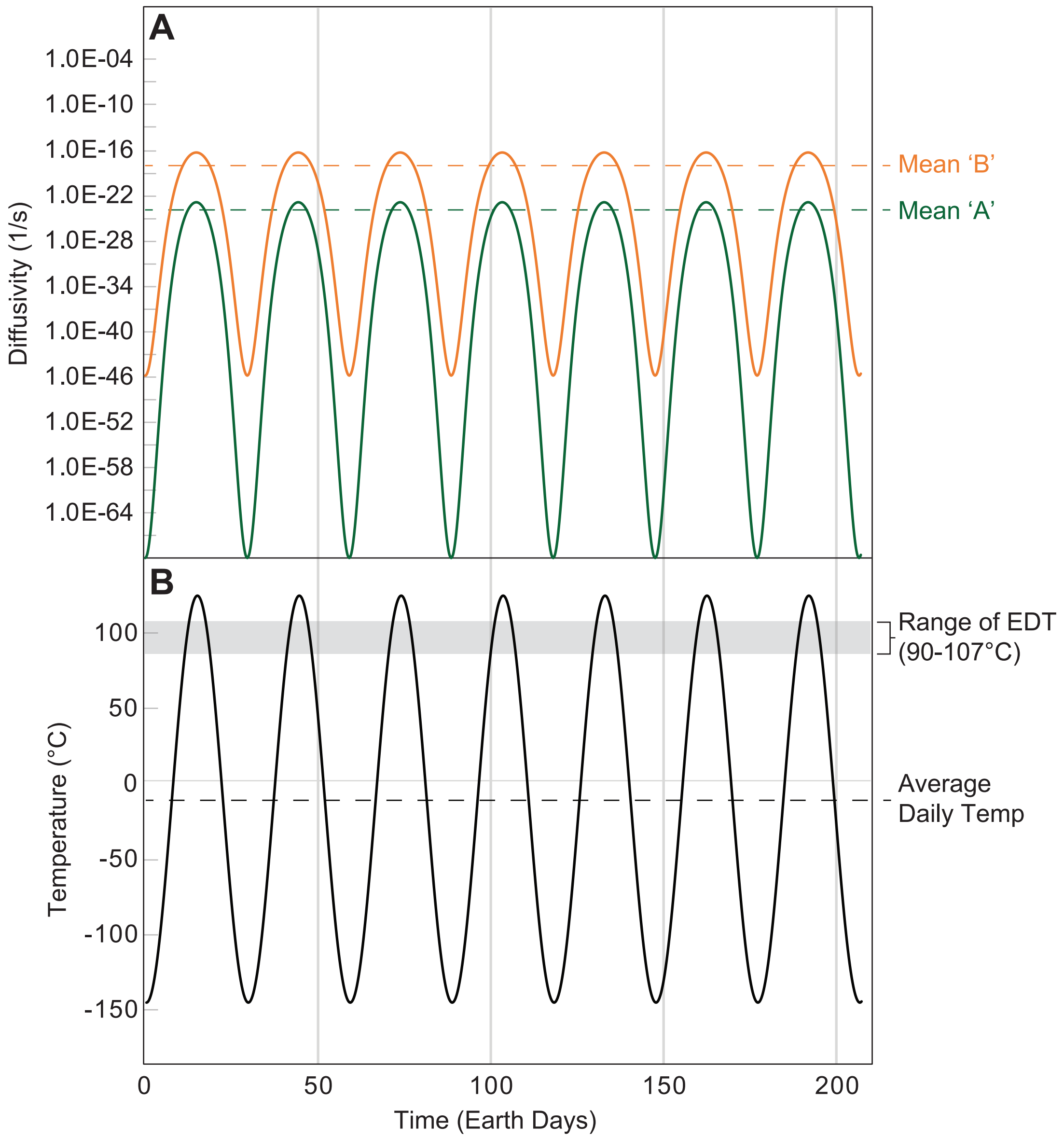

***Supplementary Figure S2.*** Schematic illustration showing the effective diffusion temperature (EDT) method (Shuster and Cassata, 2015; Tremblay et al., 2014), where a "mean diffusivity" is calculated for a sample that experiences a repeating thermal history. a) Zircon diffusivity versus time over daily temperature cycles for 2 different zircon radiation damage states based on He diffusion kinetics measured by Guenthner et al. (2013): Mean 'A': zircons RB140, BR231; Mean 'B': G3. b): daily equatorial lunar surface temperature variation plotted against the range of calculated EDT's: 90-107°C.

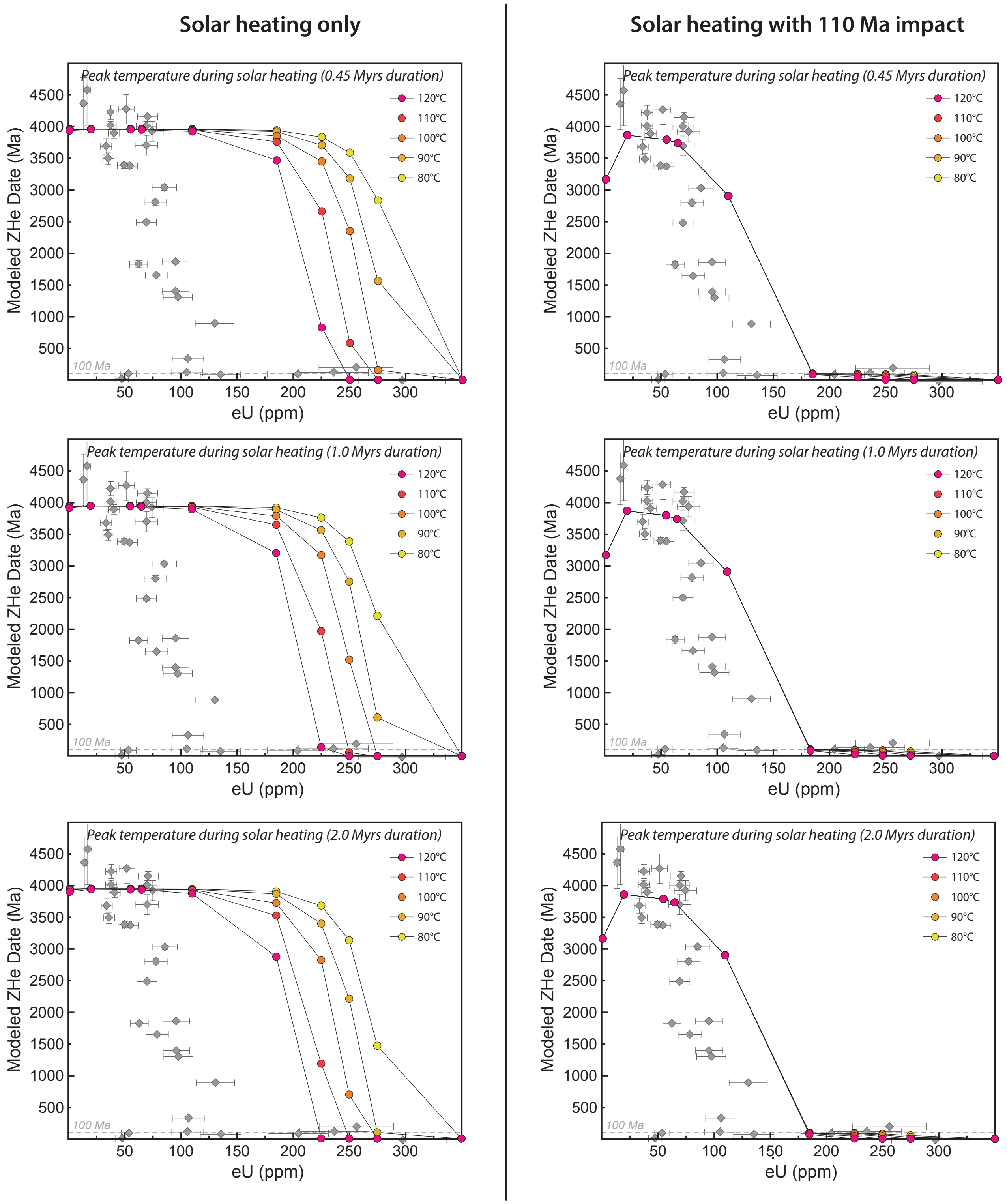

***Supplementary Figure S3.*** Date-eU results of forward modeling by HeFTy to simulate the effects of maximum daily heating of zircon grains on the lunar surface using the "EDT" approach (after Tremblay et al., 2014). Solar cycling was modeled over durations of 0.45, 1.0 and 2.0 Myrs (estimated range of at-surface exposure). 120°C represents the "maximum" daytime temperature of the outside surface of a sample during solar heating in full sunlight near the lunar equator. Models were also run for 10°C increments down to 80°C to simulate attenuated heating for zircon grains sited deeper with the sample, shading of the sample during part of the lunar day, or partial cover by regolith. Left side) Solar cycling with no impact heating following initial resetting of zircon during Imbrium impact at 3950 Ma. Right side) Solar cycling following the Imbrium impact, and a young impact event at 110 Ma (peak T = 400°C, cooling over 1 year).

## *Supplementary Figure S4*

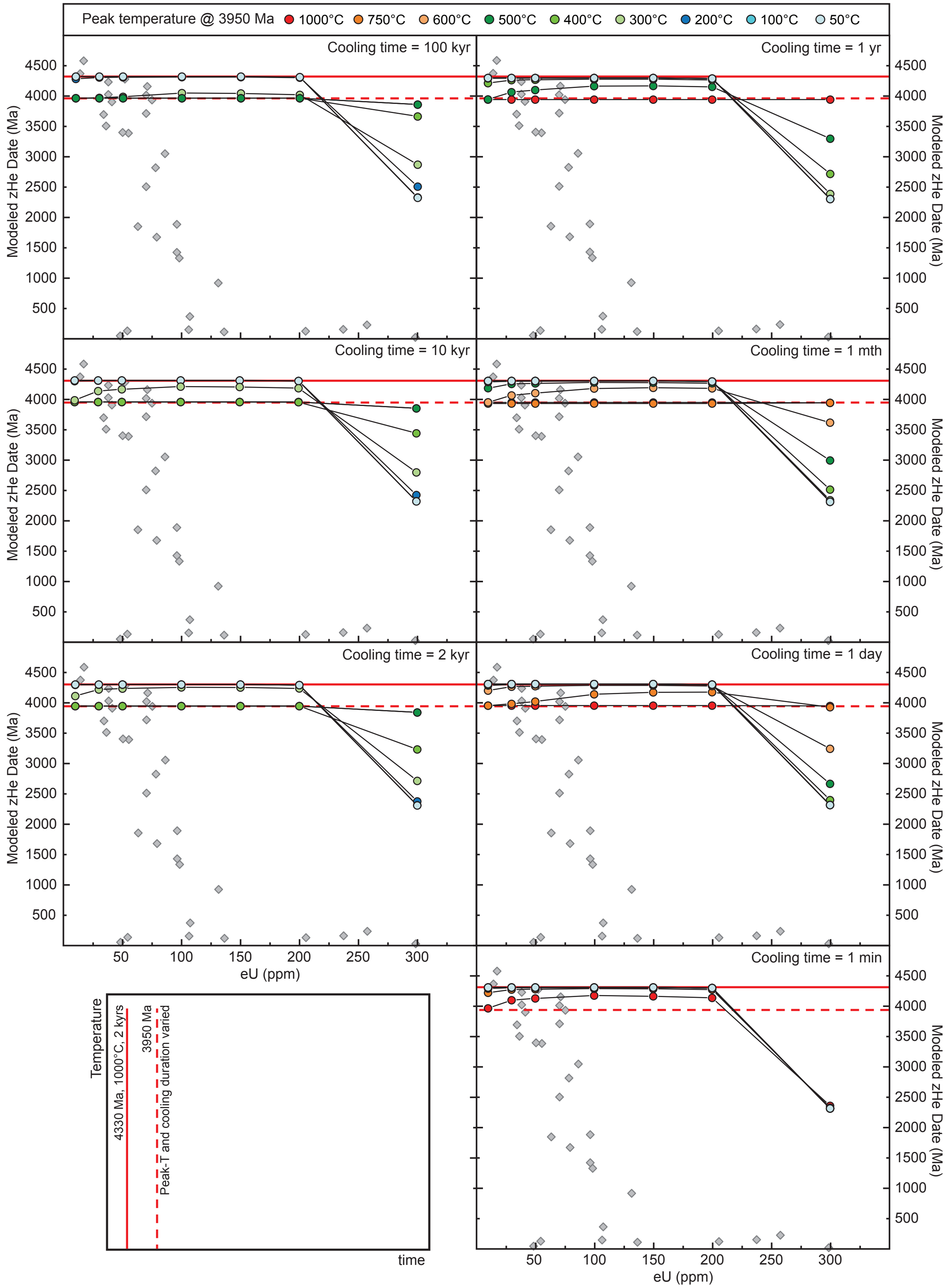

***Supplementary Figure S4.*** Date-eU plots that show measured lunar ZHe data (gray diamonds), compared with predicted date-eU patterns calculated in HeFTy forward models (Ketcham, 2005). Presented in this figure are models for a time-temperature history involving zircon grains that crystallized from a magma at 4330 Ma (the oldest U-Pb zircon population) cooling to 0°C to simulate near surface conditions by 4300 Ma (fixed for all models), and experienced a thermal event at 3950 Ma (peak-T and cooling durations varied), simulating heating within the Imbrium ejecta blanket. Red solid line indicates the starting date for the model run (4330 Ma) and red dashed line indicates the heating event (3950 Ma). The results of this series of models were used to construct Figure 3a. The t-T history is also summarized in the bottom left plot, with similar t-T summary plots provided in other model summary figures. t-T histories that include only a 3950 Ma thermal event cannot reproduce the measured lunar ZHe data.

## *Supplementary Figure S5*

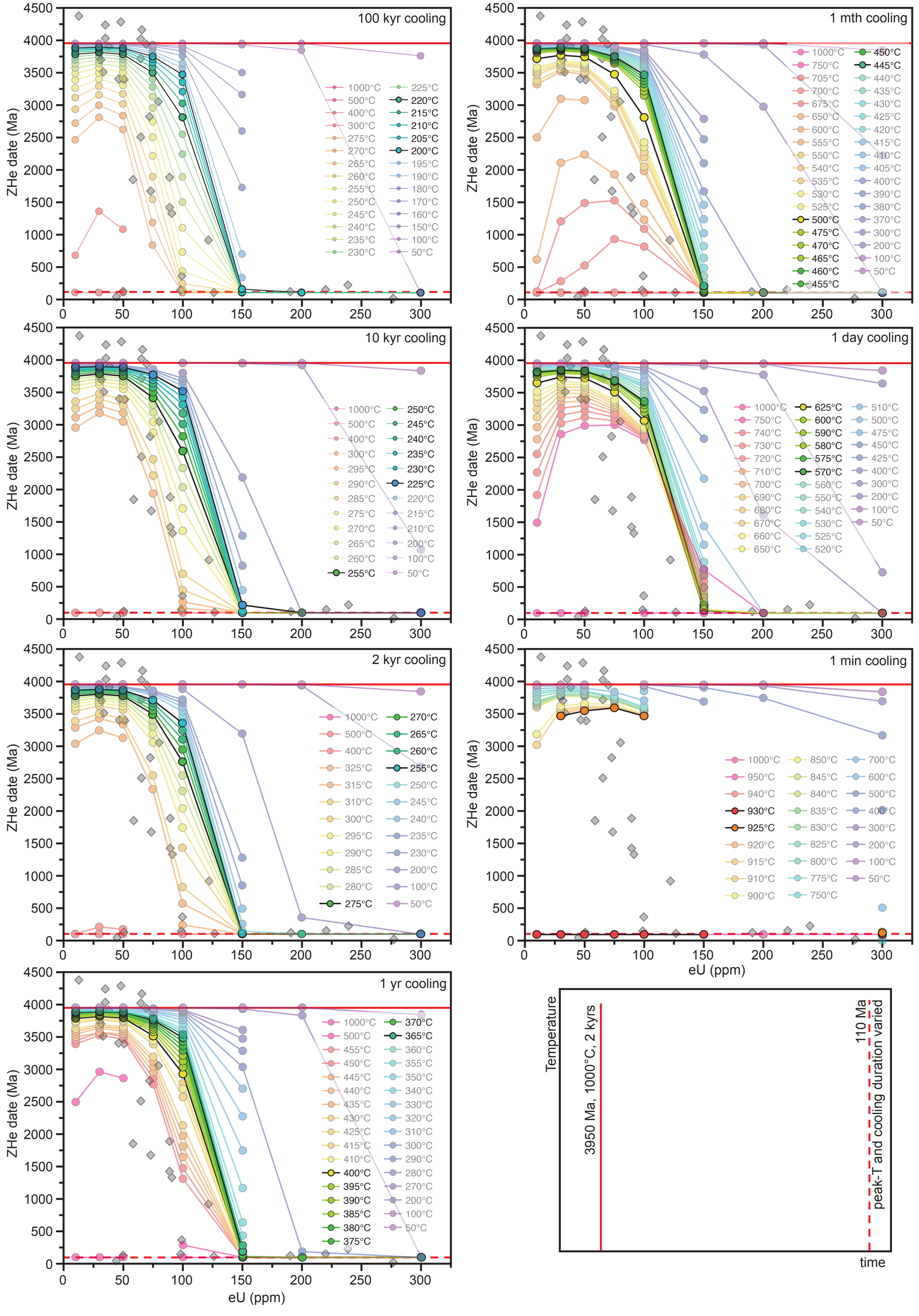

***Supplementary Figure S5.*** Date-eU plots that show measured lunar ZHe data (gray diamonds), compared with predicted date-eU patterns calculated in HeFTy forward models (Ketcham, 2005). Presented in this figure are models for a time-temperature history involving complete resetting of zircon ZHe dates during an impact at 3950 Ma (conditions fixed for all models at peak-T = 1000°C, cooling to 0°C in 2,000 years), followed by a second thermal event at 110 Ma (peak-T and cooling duration varied), simulating heating associated with a *ca.* 110 Ma impact event. Data are shaded where conditions are precluded by the measured date-eU patterns, and prominent where permitted, with results from this model series used to construct Figure 3b, c. Dashed red lines indicate the timing of thermal events. This t-T history best reproduces the measured lunar ZHe data.

# *Supplementary Figure S6*

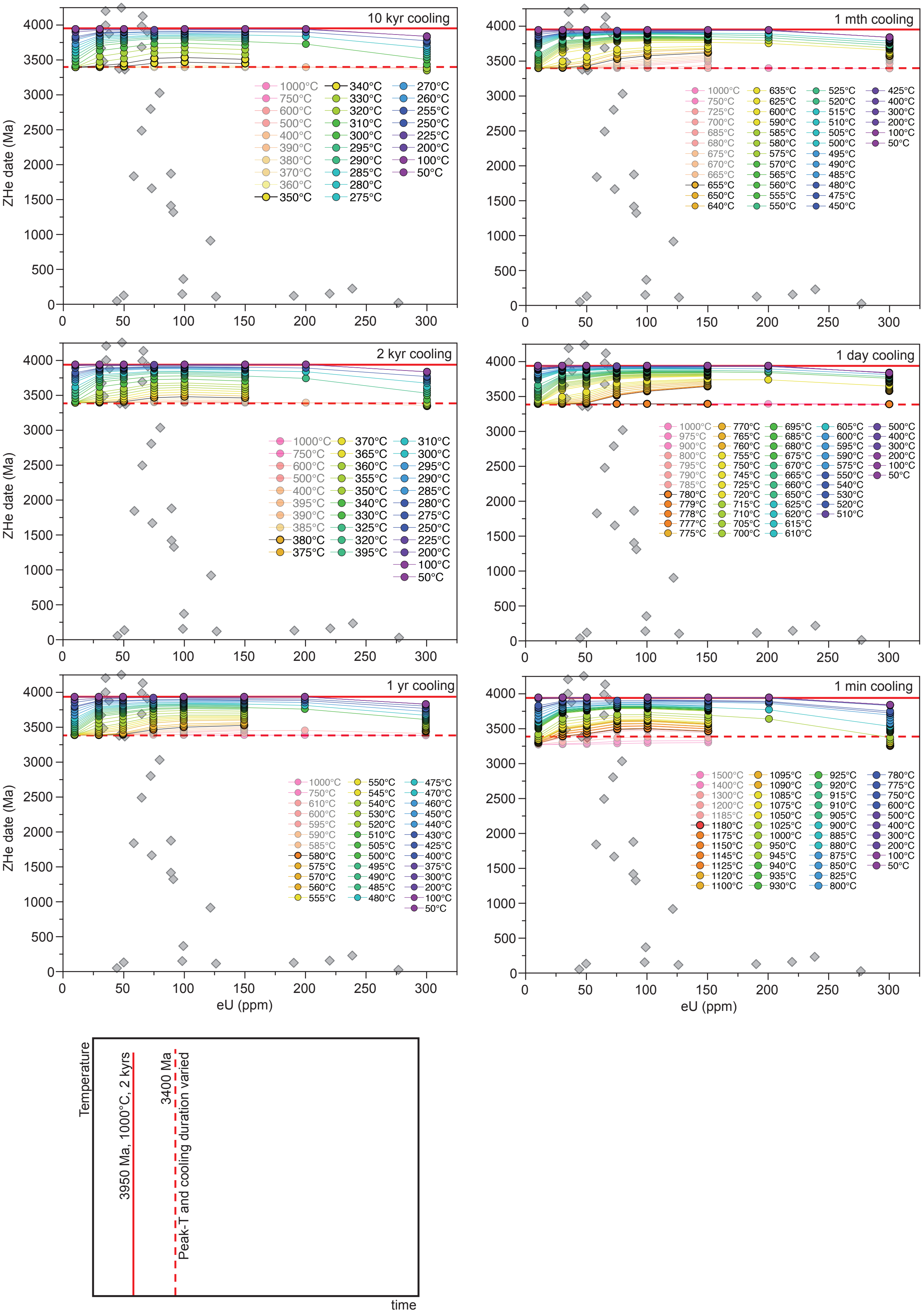

***Supplementary Figure S6.*** Date-eU plots that show measured lunar ZHe data (gray diamonds), compared with predicted date-eU patterns calculated in HeFTy forward models (Ketcham, 2005). Presented in this figure are models for a time-temperature history involving complete resetting of zircon ZHe dates at 3950 Ma (conditions fixed for all models at peak-T = 1000°C, cooling to 0°C in 2,000 years), followed by a second thermal event at 3400 Ma (peak-T and cooling duration varied), simulating hypothetical heating associated with volcanism (proposed by Pidgeon et al., 2016; Merle et al., 2017). Data are shaded where conditions are precluded by the measured date-eU patterns, and prominent where permitted, with results from this model series used to construct Figure 3c. Dashed red lines indicate the timing of thermal events. t-T histories that terminate with a 3400 Ma thermal event cannot reproduce the measured lunar ZHe data.

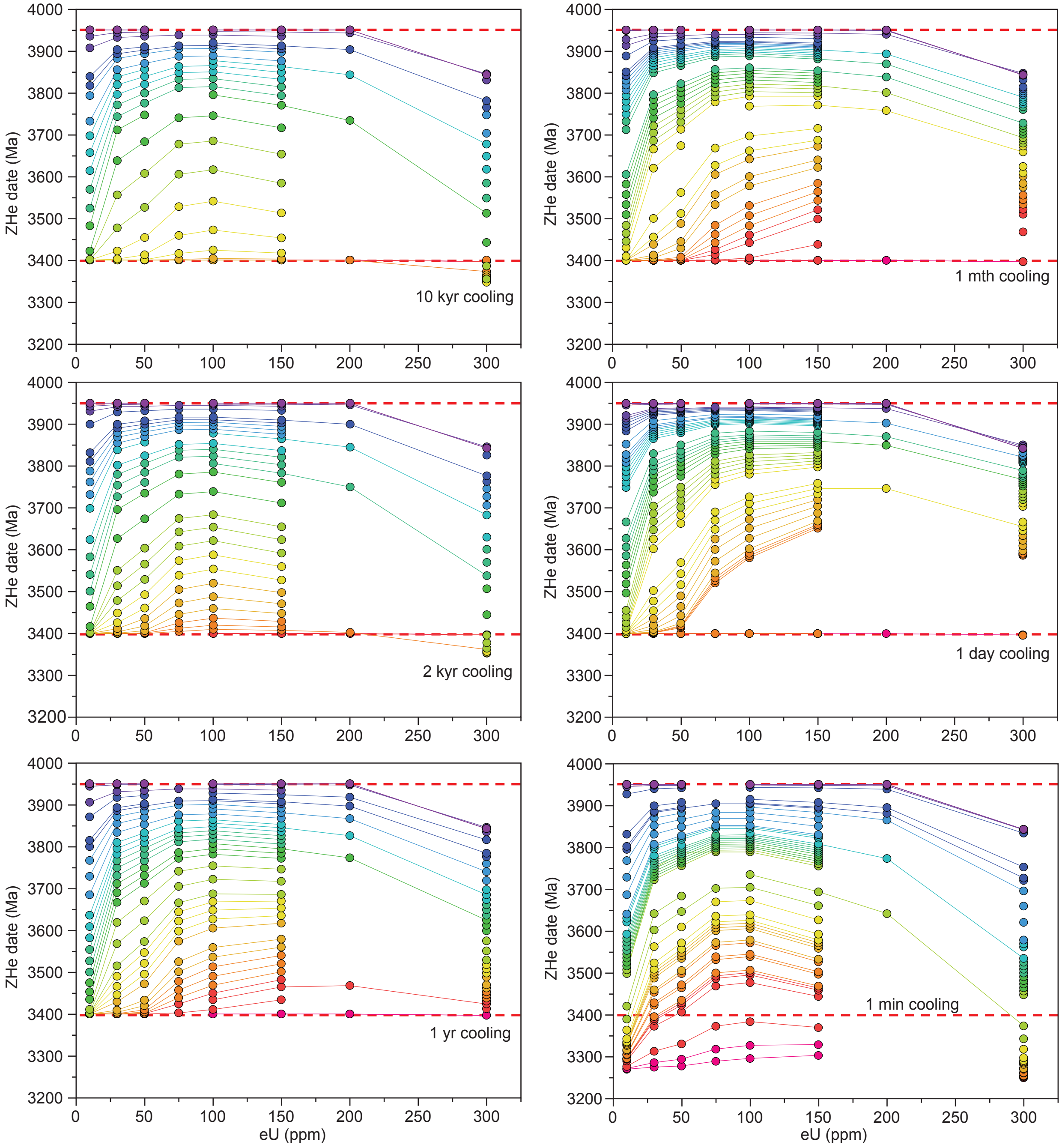

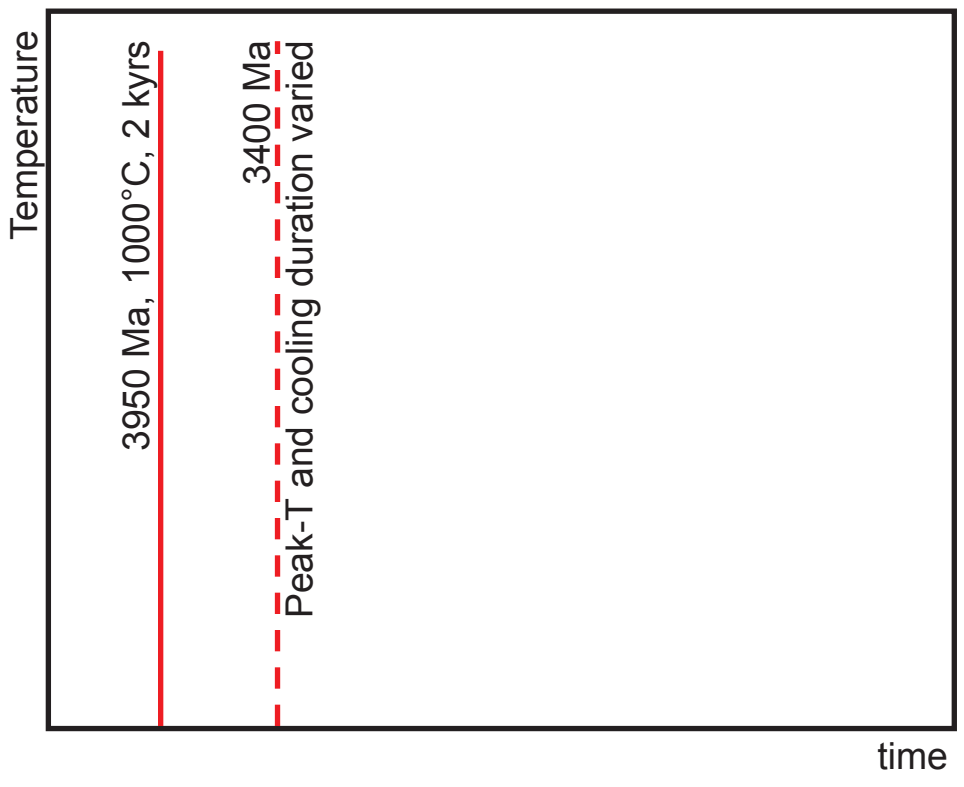

***Supplementary Figure S7.*** Reproductions of HeFTy forward models in Figure S6, but scaled to enhance dates reset in the range 3950 to 3500 Ma. Symbol legend is presented on the following page

## *Supplementary Figure S7b*

**10 kyr cooling**

1000°C, 750°C, 600°C, 500°C, 400°C, 390°C, 380°C, 370°C, 360°C, 350°C
340°C, 330°C, 320°C, 310°C, 300°C, 295°C, 290°C, 285°C, 280°C, 275°C
270°C, 260°C, 255°C, 250°C, 225°C, 200°C, 100°C, 50°C

**1 mth cooling**

1000°C, 750°C, 725°C, 700°C, 685°C, 680°C, 675°C, 670°C, 665°C, 655°C, 650°C, 640°C
635°C, 625°C, 600°C, 590°C, 585°C, 580°C, 575°C, 570°C, 565°C, 560°C, 555°C, 550°C
525°C, 520°C, 515°C, 510°C, 505°C, 500°C, 495°C, 490°C, 485°C, 480°C, 475°C, 450°C
425°C, 400°C, 300°C, 200°C, 100°C, 50°C

**2 kyr cooling**

1000°C, 750°C, 600°C, 500°C, 400°C, 395°C, 390°C, 385°C, 380°C, 375°C
370°C, 365°C, 360°C, 355°C, 350°C, 340°C, 330°C, 325°C, 320°C, 395°C
310°C, 300°C, 295°C, 290°C, 285°C, 280°C, 275°C, 250°C, 225°C, 200°C, 100°C, 50°C

**1 day cooling**

1000°C, 975°C, 900°C, 800°C, 795°C, 790°C, 785°C, 780°C, 779°C, 778°C, 777°C, 775°C, 770°C
765°C, 760°C, 755°C, 750°C, 745°C, 725°C, 720°C, 715°C, 710°C, 705°C, 700°C, 695°C, 685°C
680°C, 675°C, 670°C, 665°C, 660°C, 650°C, 625°C, 620°C, 615°C, 610°C, 605°C, 600°C, 595°C
590°C, 575°C, 550°C, 540°C, 530°C, 520°C, 510°C, 500°C, 400°C, 300°C, 200°C, 100°C, 50°C

**1 yr cooling**

1000°C, 750°C, 610°C, 600°C, 595°C, 590°C, 585°C, 580°C, 575°C, 570°C, 560°C, 555°C
550°C, 545°C, 540°C, 530°C, 520°C, 510°C, 505°C, 500°C, 495°C, 490°C, 485°C, 480°C
475°C, 470°C, 460°C, 450°C, 440°C, 430°C, 425°C, 400°C, 375°C, 300°C, 200°C, 100°C, 50°C

**1 min cooling**

1500°C, 1400°C, 1300°C, 1200°C, 1185°C, 1180°C, 1175°C, 1150°C, 1145°C, 1125°C, 1120°C, 1100°C
1095°C, 1090°C, 1085°C, 1075°C, 1050°C, 1025°C, 1000°C, 950°C, 945°C, 940°C, 935°C, 930°C
925°C, 920°C, 915°C, 910°C, 905°C, 900°C, 885°C, 880°C, 875°C, 850°C, 825°C, 800°C
780°C, 775°C, 750°C, 600°C, 500°C, 400°C, 300°C, 200°C, 100°C, 50°C

## *Supplementary Figure S8*

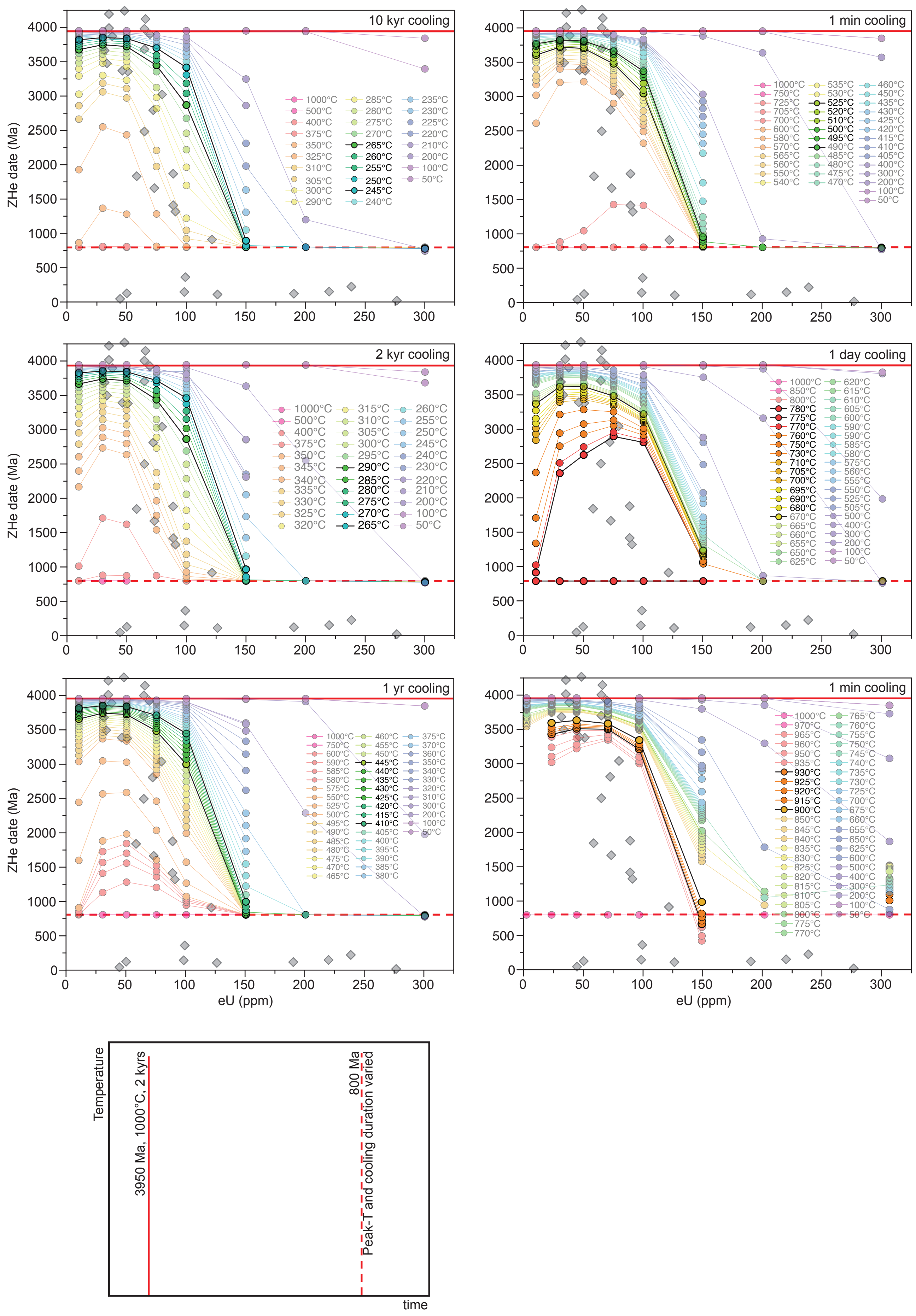

***Supplementary Figure S8.*** Date-eU plots that show measured lunar ZHe data (shaded pale blue diamonds), compared with predicted date-eU patterns calculated in HeFTy forward models (Ketcham, 2005). Presented in this figure are models for a time-temperature history involving complete resetting of zircon ZHe dates at 3950 Ma (conditions fixed for all models at peak-T = 1000°C, cooling to 0°C in 2,000 years), followed by a second thermal event at 800 Ma (peak-T and cooling duration varied), simulating hypothetical heating associated with a Copernicus-aged impact. Data are shaded where conditions are precluded by the measured date-eU patterns, and prominent where permitted. Dashed red lines indicate the timing of thermal events. t-T histories that terminate with an 800 Ma thermal event cannot reproduce the measured lunar ZHe data.

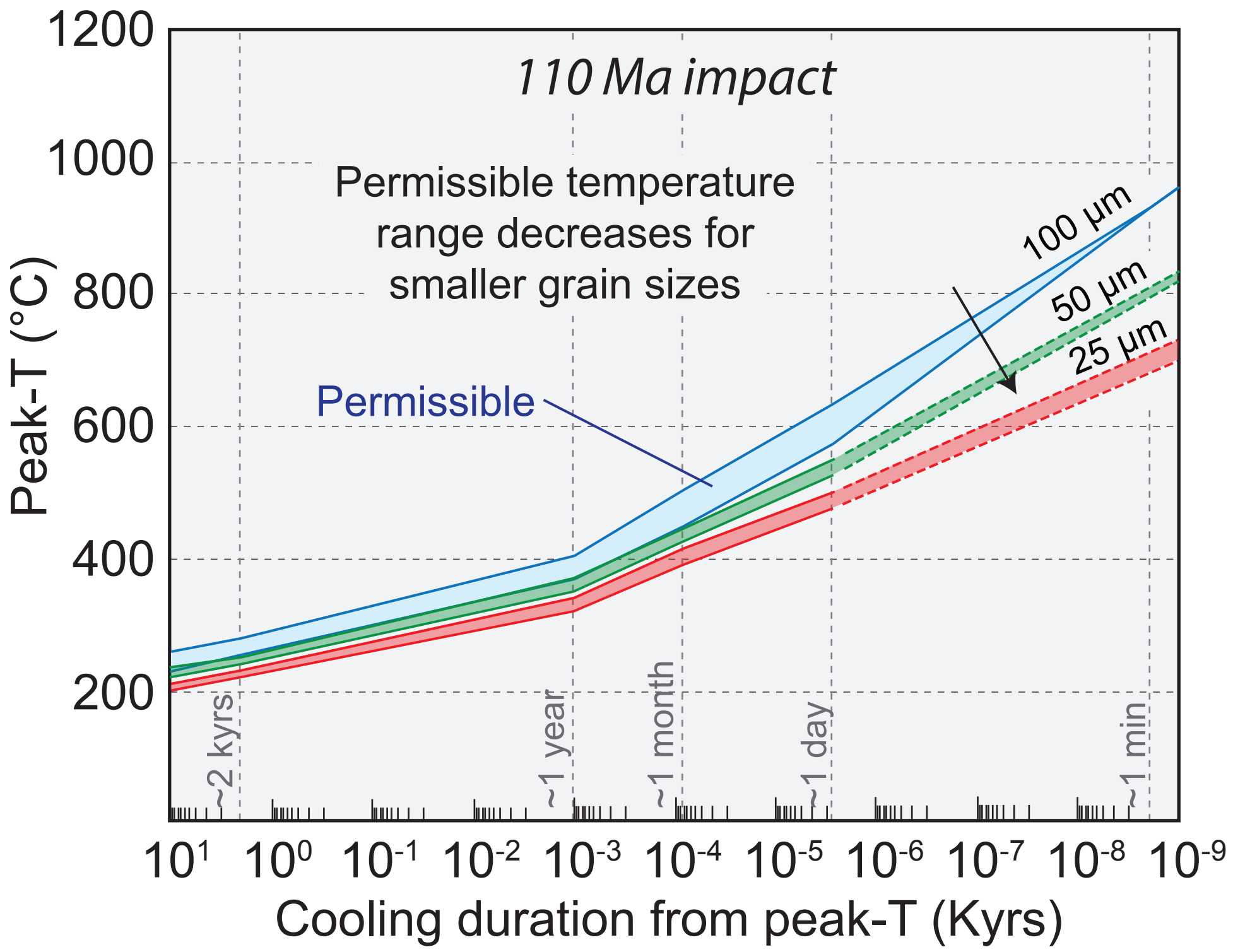

***Supplementary Figure S9***. Constraints from HeFTy models for temperatures permissible or precluded during an impact event at 110 Ma following accumulation of radiation damage since 3950 Ma. The plot illustrates the projected differences in permissible temperatures for this event based on grain sizes 100, 50 and 25 µm.

*Supplementary Table S1. Zircon (U-Th)/He data vs $^{207}Pb/^{206}Pb$ age for lunar impact-melt breccia sample 14311.*

| Sample & Grain | U (ppm) | Th (ppm) | Sm (ppm) | eU (ppm) | ± (ppm) | $^{4}He$ Meas (ncc) | GCR Corr. $^{4}He$ (ncc) | ± | GCR-corr Date (Ma) | ± (Ma) | $^{207}Pb/^{206}Pb$ Age Ma | ± Ma | Interpreted U-Pb population* |
|---|---|---|---|---|---|---|---|---|---|---|---|---|---|
| **14311.20** | | | | | | | | | | | | | |
| 20-3_z2 | 190.7 | 59.0 | 0.0 | 205 | 27 | 3.2 | 3.1 | 0.00 | 109 | 2 | 4215 | 20 | 4.25 Ga |
| 20-3_z5 | 90.2 | 34.7 | 4.8 | 98 | 13 | 24.6 | 24.5 | 0.03 | 1318 | 24 | 4352 | 116 | 4.33 Ga |
| 20-3_z9 | 109.3 | 115.3 | 4.0 | 136 | 18 | 3.0 | 2.9 | 0.00 | 96 | 3 | 4291 | 4 | Partial reset |
| 20-3_z10 | 114.3 | 71.1 | 4.2 | 131 | 17 | 102.5 | 102.1 | 0.07 | 906 | 24 | 4332 | 7 | 4.33 Ga |
| 20-3_z13 | 82.5 | 56.9 | 1.5 | 96 | 12 | 88.4 | 88.1 | 0.09 | 1412 | 16 | 4225 | 9 | 4.25 Ga |
| 20-3_z14 | 91.5 | 64.1 | 1.8 | 107 | 14 | 26.1 | 25.7 | 0.02 | 350 | 3 | 4333 | 9 | 4.33 Ga |
| 20-3_z17 | 29.5 | 28.8 | 0.3 | 36 | 5 | 67.1 | 66.9 | 0.06 | 3502 | 88 | 4216 | 15 | 4.25 Ga |
| 20-4_z1 | 64.7 | 44.6 | 16.3 | 75 | 10 | 47.6 | 47.6 | 0.03 | 3933 | 157 | 4250 | 8 | 4.25 Ga |
| 20-4_z2 | 12.0 | 7.6 | 16.7 | 14 | 2 | 13.1 | 13.0 | 0.01 | 4369 | 407 | 3964 | 28 | 3.95 Ga |
| 20-4_z7$_{h}$ | 46.0 | 24.4 | 8.0 | 52 | 7 | 72.0 | 71.9 | 0.03 | 4278 | 230 | 4254 | 15 | 4.25 Ga |
| 20-4_z8$^{h}$ | 240.3 | 245.6 | 6.5 | 298 | 39 | 0.6 | 0.5 | 0.00 | 6 | 2 | 4236 | 11 | 4.25 Ga |
| 20-5_z3 | 67.3 | 51.7 | 0.0 | 79 | 10 | 67.8 | 67.6 | 0.06 | 1665 | 24 | 4331 | 13 | 4.33 Ga |
| 20-5_z5 | 47.7 | 29.5 | 7.6 | 55 | 7 | 66.5 | 66.4 | 0.06 | 3383 | 36 | 4340 | 12 | 4.33 Ga |
| **14311.50** | | | | | | | | | | | | | |
| 50-3_z4 | 35.6 | 21.0 | 4.7 | 41 | 5 | 139.4 | 139.1 | 0.44 | 3902 | 78 | 4279 | 18 | 4.25 Ga |
| 50-3_z6 | 60.4 | 39.2 | 9.8 | 70 | 9 | 120.5 | 120.4 | 0.21 | 4011 | 78 | 4242 | 10 | 4.25 Ga |
| 50-3_z7 | 91.6 | 60.3 | 4.0 | 106 | 14 | 3.2 | 3.1 | 0.00 | 132 | 4 | 4278 | 8 | 4.25 Ga |
| 50-3_z10 | 211.6 | 107.4 | 3.8 | 237 | 31 | 19.1 | 18.8 | 0.01 | 138 | 2 | 4211 | 6 | 4.25 Ga |
| 50-3_z11 | 60.2 | 45.8 | 4.9 | 71 | 9 | 329.7 | 329.4 | 0.32 | 4157 | 74 | 4240 | 8 | 4.25 Ga |
| 50-3_z12 | 29.4 | 18.1 | 4.1 | 34 | 5 | 66.5 | 66.3 | 0.08 | 3692 | 122 | 4080 | 55 | Partial reset |
| **14311.60** | | | | | | | | | | | | | |
| 60-2_z4 | 75.6 | 45.6 | 0.0 | 86 | 11 | 88.9 | 88.7 | 0.02 | 3045 | 45 | 4246 | 13 | 4.25 Ga |
| 60-4_z6 | 70.5 | 31.8 | 9.1 | 78 | 10 | 102.5 | 102.3 | 0.06 | 2812 | 50 | 4111 | 24 | Partial reset |
| 60-4_z7 | 62.4 | 32.5 | 53.7 | 70 | 10 | 53.7 | 53.6 | 0.03 | 3710 | 158 | 4252 | 10 | 4.25 Ga |
| 60-4_z9 | 64.5 | 21.5 | 0.0 | 70 | 9 | 142.4 | 142.1 | 0.08 | 2499 | 28 | 4252 | 8 | 4.25 Ga |
| 60-4_z10 | 44.7 | 20.7 | 0.0 | 50 | 6 | 152.1 | 151.8 | 0.13 | 3395 | 49 | 3971 | 10 | 3.95 Ga |
| 60-5_z1 | 14.8 | 9.1 | 0.0 | 17 | 3 | 20.8 | 20.8 | 0.03 | 4580 | 555 | 3960 | 18 | 3.95 Ga |
| 60-5_z2 | 87.9 | 34.2 | 5.5 | 96 | 12 | 94.5 | 94.3 | 0.41 | 1875 | 22 | 3937 | 9 | 3.95 Ga |
| 60-5_z3 | 32.6 | 22.4 | 1.0 | 38 | 5 | 165.5 | 165.1 | 0.79 | 4023 | 62 | 4305 | 12 | 4.33 Ga |
| 60-5_z4 | 231.3 | 107.8 | 10.9 | 257 | 33 | 6.4 | 6.3 | 0.01 | 212 | 2 | 4382 | 22 | 4.33 Ga |
| 60-5_z5 | 56.5 | 26.8 | 27.9 | 63 | 8 | 17.5 | 17.5 | 0.02 | 1838 | 54 | 3971 | 15 | 3.95 Ga |
| 60-5_z7 | 43.6 | 18.6 | 0.0 | 48 | 6 | 0.3 | 0.2 | 0.00 | 31 | 11 | 4119 | 9 | Partial reset |
| 60-5_z9 | 46.7 | 32.5 | 1.9 | 54 | 7 | 1.3 | 1.3 | 0.39 | 115 | 2 | 4344 | 7 | 4.33 Ga |
| 60-5_z11 | 33.0 | 21.4 | 0.0 | 38 | 5 | 85.1 | 84.9 | 0.79 | 4231 | 111 | 4272 | 29 | 4.25 Ga |

*U/Pb populations (baed on Hopkins & Mojzsis, 2015):

ca. 4.33 Ga: igneous zircon, lunar crust formation

ca. 4.25 Ga: igneous zircon or growth in large impact melt sheet

ca. 3.95 Ga: impact shock resetting and neoblastic zircon growth in impact melt

***Supplementary Table S2. Data for grains with ZHe dates <660 Ma, with cosmogenic $^{4}$He recalcuated for measured ZHe data***

| Sample & Grain | Vol.[a] (mm$^3$ x10$^{-3}$) | Mass[b] (µg) | Spherical radius from mass (µm) | U (ppm) | Th (ppm) | Sm (ppm) | eU[c] (ppm) | ±[d] (ppm) | $^4$He (nmol/g) | ± | $^4$He Meas (ncc) | GCR[e] accum. time (Ma) | Cosmo-genic $^4$He[f] (ncc) | % $^4$He$_{cos}$ of meas. $^4$He | GCR Corr. $^4$He (ncc) | ± | GCR-corr Date (Ma) | ± (Ma) | Date diff.[g] (Ma) |
|---|---|---|---|---|---|---|---|---|---|---|---|---|---|---|---|---|---|---|---|
| **14311.20** | | | | | | | | | | | | | | | | | | | |
| 20-3_z2 | 0.25 | 1.15 | 39 | 190.7 | 59.0 | 0.0 | 205 | 27 | 123.4 | 0.2 | 3.2 | 110 | 0.010 | 0.32 | 3.2 | 0.00 | 110.8 | 1.9 | 2.2 |
| 20-3_z9 | 0.39 | 1.83 | 45 | 109.3 | 115.3 | 4.0 | 136 | 18 | 73.4 | 0.1 | 3.0 | 99 | 0.018 | 0.59 | 3.0 | 0.00 | 99.0 | 2.7 | 3.3 |
| 20-3_z14 | 1.19 | 5.52 | 66 | 91.5 | 64.1 | 1.8 | 107 | 14 | 209.6 | 0.2 | 26.1 | 353 | 0.193 | 0.74 | 25.9 | 0.02 | 353.7 | 3.8 | 3.4 |
| 20-4_z8 | | 2.42 | 50 | 240.3 | 245.6 | 6.5 | 298 | 39 | 11.9 | 0.0 | 0.6 | 7 | 0.002 | 0.27 | 0.6 | 0.00 | 7.4 | 0.1 | 1.8 |
| **14311.50** | | | | | | | | | | | | | | | | | | | |
| 50-3_z7 | 0.39 | 1.81 | 45 | 91.6 | 60.3 | 4.0 | 106 | 14 | 78.6 | 0.1 | 3.2 | 136 | 0.024 | 0.76 | 3.2 | 0.00 | 136.2 | 1.8 | 3.9 |
| 50-3_z10 | 1.01 | 4.70 | 62 | 211.6 | 107.4 | 3.8 | 237 | 31 | 180.4 | 0.1 | 19.1 | 139 | 0.065 | 0.34 | 19.0 | 0.01 | 139.5 | 3.1 | 1.8 |
| **14311.60** | | | | | | | | | | | | | | | | | | | |
| 60-5_z4 | 0.20 | 0.94 | 36 | 231.3 | 107.8 | 10.9 | 257 | 33 | 301.5 | 0.4 | 6.4 | 213 | 0.020 | 0.31 | 6.4 | 0.01 | 213.9 | 2.8 | 2.1 |
| 60-5_z7 | 0.24 | 1.11 | 38 | 43.6 | 18.6 | 0.0 | 48 | 6 | 10.9 | 0.1 | 0.3 | 42 | 0.005 | 1.67 | 0.3 | 0.00 | 42.1 | 1.4 | 10.6 |
| 60-5_z9 | 0.36 | 1.68 | 44 | 46.7 | 32.5 | 1.9 | 54 | 7 | 34.7 | 10.4 | 1.3 | 117 | 0.005 | 0.36 | 1.3 | 0.39 | 117.0 | 35.3 | 2.4 |

[a] Volume measured by HR-XCT

[b] Mass calculated using density of pristine zircon (4.65 g/cm$^3$)

[c] eU - effective uranium concentration, combines U and Th weighted by their alpha productivity, computed as [U] + 0.235 * [Th]

[d] uncertainty includes analytical error and uncertainty involving estimated density of pristine versus metamict zircon

[e] GCR - Galactic Cosmic Rays; $^4$He may be produced due to iteractions with GCRs and is referred to here as cosmogenic $^4$He ($^4$He$_{cos}$)

[f] Cosmogenic $^4$He calculated using maximum production rates for $^4$He from (31), based on estimated composition of the breccia matrix

[g] Age difference between dates calculated using the maximum exposure age (c. 661 Ma, 19) and ZHe date for those grains where ZHe < 661 Ma.

[h] Volume estimate based on grain dimensions

*Supplementary Table S3. Cosmogenic correction data for lunar zircons from 14311.*

| | | | | Zircon P4 Max = 6.67e-08 | | | Matrix P4 Max = 9.91e-08 | | |
|---|---|---|---|---|---|---|---|---|---|
| | | | | $cm^3$ STP/g Ma[d] | | | $cm^3$ STP/g Ma[e] | | |
| | Mass (g) | Exp. age[b] | $^4$He ncc (meas)[c] | $^4$He cos (ncc)[d] | Corr. $^4$He (ncc) | %$^4$He cos | $^4$He cos (ncc) | Corr. $^4$He (ncc) | %$^4$He cos |
| **20-3_z2** | 0.0000009 | 661 | 3.2 | 0.041 | 3.1 | 1.3 | 0.061 | 3.1 | 1.9 |
| **20-3_z5** | 0.0000014 | 661 | 24.6 | 0.061 | 24.5 | 0.2 | 0.090 | 24.5 | 0.4 |
| **20-3_z9** | 0.0000018 | 661 | 3.0 | 0.081 | 3.0 | 2.7 | 0.120 | 2.9 | 4.0 |
| **20-3_z10** | 0.0000065 | 661 | 102.5 | 0.289 | 102.2 | 0.3 | 0.429 | 102.1 | 0.4 |
| **20-3_z13** | 0.0000047 | 661 | 88.4 | 0.208 | 88.2 | 0.2 | 0.309 | 88.1 | 0.3 |
| **20-3_z14** | 0.0000055 | 661 | 26.1 | 0.243 | 25.9 | 0.9 | 0.361 | 25.7 | 1.4 |
| **20-3_z17** | 0.0000029 | 661 | 67.1 | 0.126 | 67.0 | 0.2 | 0.188 | 66.9 | 0.3 |
| **20-4_z1** | 0.0000008 | 661 | 47.6 | 0.035 | 47.6 | 0.1 | 0.051 | 47.6 | 0.1 |
| **20-4_z2** | 0.0000009 | 661 | 13.1 | 0.041 | 13.1 | 0.3 | 0.061 | 13.0 | 0.5 |
| ***20-4_z7[a]*** | *0.0000014* | 661 | 72.0 | 0.063 | 71.9 | 0.1 | 0.094 | 71.9 | 0.1 |
| ***20-4_z8[a]*** | *0.0000024* | 661 | 0.6 | 0.107 | 0.5 | 16.5 | 0.159 | 0.5 | 24.5 |
| **20-5_z3** | 0.0000036 | 661 | 68.0 | 0.159 | 67.9 | 0.2 | 0.236 | 67.8 | 0.3 |
| **20-5_z5** | 0.0000020 | 661 | 66.6 | 0.087 | 66.6 | 0.1 | 0.128 | 66.5 | 0.2 |
| **50-3_z4** | 0.0000043 | 661 | 139.4 | 0.189 | 139.2 | 0.1 | 0.281 | 139.1 | 0.2 |
| **50-3_z6** | 0.0000021 | 661 | 120.5 | 0.091 | 120.4 | 0.1 | 0.134 | 120.4 | 0.1 |
| **50-3_z7** | 0.0000018 | 661 | 3.2 | 0.080 | 3.1 | 2.5 | 0.119 | 3.1 | 3.7 |
| **50-3_z10** | 0.0000047 | 661 | 19.1 | 0.207 | 18.9 | 1.1 | 0.308 | 18.8 | 1.6 |
| **50-3_z11** | 0.0000052 | 661 | 329.7 | 0.228 | 329.5 | 0.1 | 0.338 | 329.4 | 0.1 |
| **50-3_z12** | 0.0000027 | 661 | 66.5 | 0.120 | 66.4 | 0.2 | 0.179 | 66.3 | 0.3 |
| **60-2_z4** | 0.0000020 | 661 | 89.0 | 0.086 | 88.9 | 0.1 | 0.128 | 88.9 | 0.1 |
| **60-4_z6** | 0.0000028 | 661 | 102.5 | 0.123 | 102.3 | 0.1 | 0.183 | 102.3 | 0.2 |
| **60-4_z7** | 0.0000010 | 661 | 53.7 | 0.046 | 53.6 | 0.1 | 0.068 | 53.6 | 0.1 |
| **60-4_z9** | 0.0000049 | 661 | 152.5 | 0.215 | 152.2 | 0.1 | 0.319 | 152.1 | 0.2 |
| **60-4_z10** | 0.0000051 | 661 | 142.8 | 0.226 | 142.6 | 0.2 | 0.335 | 142.4 | 0.2 |
| **60-5_z1** | 0.0000011 | 661 | 20.8 | 0.048 | 20.8 | 0.2 | 0.071 | 20.8 | 0.3 |
| **60-5_z2** | 0.0000036 | 661 | 94.5 | 0.158 | 94.3 | 0.2 | 0.234 | 94.3 | 0.2 |
| **60-5_z3** | 0.0000052 | 661 | 165.5 | 0.228 | 165.2 | 0.1 | 0.339 | 165.1 | 0.2 |
| **60-5_z4** | 0.0000009 | 661 | 6.4 | 0.042 | 6.3 | 0.7 | 0.062 | 6.3 | 1.0 |
| **60-5_z5** | 0.0000010 | 661 | 17.5 | 0.046 | 17.5 | 0.3 | 0.068 | 17.5 | 0.4 |
| **60-5_z7** | 0.0000011 | 661 | 0.3 | 0.049 | 0.2 | 17.7 | 0.072 | 0.2 | 26.3 |
| **60-5_z9** | 0.0000004 | 661 | 1.3 | 0.018 | 1.3 | 1.4 | 0.027 | 1.3 | 2.1 |
| **60-5_z11** | 0.0000024 | 661 | 85.1 | 0.105 | 85.0 | 0.1 | 0.156 | 84.9 | 0.2 |

***Grains where ZHe was less than published exposure age (c. 661 Ma)[g]***

| | | | | Zircon P4 Max = 6.67e-08[d] | | | Matrix P4 Max = 9.91e-08[e] | | | | | |
|---|---|---|---|---|---|---|---|---|---|---|---|---|
| | Mass (g) | GCR exposure time (Ma)[h] | $^4$He ncc (meas)[c] | $^4$He cos (ncc) | Corr. $^4$He (ncc) | %$^4$He cos | $^4$He cos (ncc) | Corr. $^4$He (ncc) | %$^4$He cos | ZHe date GCR-661 Ma[i] | ZHe date GCR-Meas. Date[j] | Diff. (Ma) |
| **20-3_z2** | 0.0000011 | 110.4 | 3.18 | 0.008 | 3.2 | 0.3 | 0.013 | 3.17 | 0.4 | 108.6 | 110.4 | 1.83 |
| **20-3_z9** | 0.0000018 | 99.0 | 3.04 | 0.012 | 3.0 | 0.4 | 0.018 | 3.02 | 0.6 | 95.9 | 99.0 | 3.05 |
| **20-3_z14** | 0.0000055 | 352.6 | 26.11 | 0.130 | 26.0 | 0.5 | 0.193 | 25.92 | 0.7 | 351.5 | 352.6 | 1.10 |
| ***20-4_z8[a]*** | *0.0000024* | 7.4 | 0.65 | 0.001 | 0.6 | 0.2 | 0.002 | 0.65 | 0.3 | 5.6 | 7.4 | 1.78 |
| **50-3_z7** | 0.0000018 | 136.2 | 3.34 | 0.016 | 3.3 | 0.5 | 0.024 | 3.32 | 0.7 | 137.2 | 136.2 | -1.02 |
| **50-3_z10** | 0.0000047 | 139.5 | 19.08 | 0.044 | 19.0 | 0.2 | 0.065 | 19.01 | 0.3 | 137.8 | 139.5 | 1.66 |
| **60-5_z4** | 0.0000009 | 213.1 | 6.39 | 0.013 | 6.4 | 0.2 | 0.020 | 6.37 | 0.3 | 212.5 | 213.1 | 0.62 |
| **60-5_z7** | 0.0000011 | 41.9 | 0.28 | 0.003 | 0.3 | 1.1 | 0.005 | 0.27 | 1.7 | 31.6 | 41.9 | 10.35 |
| **60-5_z9** | 0.0000004 | 116.6 | 1.31 | 0.003 | 1.3 | 0.2 | 0.005 | 1.31 | 0.4 | 107.6 | 116.6 | 8.91 |

[a] mass estimate by measurement of grain dimensions, not HRXCT

[b] Exposure age based on $^{81}$Kr-Kr (Crozaz et al., 1972)

[c] Measured ZHe dates (see Table 1 for full details)

[d] P4 Max = maximum production rate of $^4$He from zircon, using values from Leya et al. (2004)

[e] P4 Max = maximum production rate of $^4$He for the impact-melt breccia matrix, using values from Leya et al. (2004) corrected for matrix composition based on breccia sample 14311,67 (Scoon, 1972).

[f] $^4He_{COS}$ = total cosmogenic production of $^4$He

[g] $^4He_{COS}$ was calculated using the measured ZHe date (based on assumption that $^4He_{COS}$ would be lost proportionally with $^4He_{RAD}$)

[h] Accumulation time used for calculation of cosmogenic $^4$He; this age was calculated by iterating $^4He_{COS}$ values to calculate new corrected dates.

[i] Original ZHe date, corrected for $^4He_{COS}$ using exposure age (661 Ma)

[j] ZHe date corrected for $^4He_{COS}$ using measured age

*Supplementary Table S4. Output from Effective Diffusion Temperature calculations used in HeFTy forward modeling of solar cycling*

| | From Guenthner et al. (2013) | | | Estimated eU for equivalent α-dose after 3.95 Ga[1] | | 120°C Max Solar Heating | | 110°C Max Solar Heating | | 100°C Max Solar Heating | | 90°C Max Solar Heating | | 80°C Max Solar Heating | |
|---|---|---|---|---|---|---|---|---|---|---|---|---|---|---|---|
| | Published α-dose ($x10^{16}$) | Published Do (cm2/s) | Published Ea (KJ/mol) | eU (ppm) | Calc. a-dose @ 3950 Ma ($x10^{16}$) | Mean Diff | Calculated EDT[2] | Mean Diff | Calculated EDT | Mean Diff | Calculated EDT | Mean Diff | Calculated EDT | Mean Diff | Calculated EDT |
| **Mud Tank** | **1.22** | 110.50 | 167.93 | 1 | 2.2 | 9.718E-22 | 107 | 1.362E-22 | 95 | 3.376E-23 | 82 | 7.767E-24 | 76 | 1.650E-24 | 66 |
| **RB140** | **46.7** | 0.2011 | 166.49 | 20 | 45 | 2.744E-24 | 108 | 3.913E-25 | 95 | 9.815E-26 | 83 | 2.287E-26 | 74 | 4.923E-27 | 66 |
| **BR231** | **121** | 0.2304 | 169.75 | 55 | 120 | 1.163E-24 | 107 | 1.596E-25 | 95 | 3.894E-26 | 83 | 8.816E-27 | 75 | 1.841E-27 | 66 |
| **M127** | **148** | 0.0265 | 161.58 | 65 | 145 | 1.617E-24 | 107 | 2.444E-25 | 92 | 6.390E-26 | 83 | 1.555E-26 | 75 | 3.506E-27 | 67 |
| **TH62Z** | **250** | 0.0170 | 145.96 | 110 | 245 | 1.222E-22 | 105 | 2.222E-23 | 91 | 6.624E-24 | 83 | 1.852E-24 | 75 | 4.834E-25 | 65 |
| **G3** | **404** | 0.0042 | 106.53 | 185 | 404 | 5.240E-18 | 100 | 1.519E-18 | 88 | 6.314E-19 | 80 | 2.506E-19 | 68 | 9.471E-20 | 60 |
| **Model-zrc-1** | *EDT values through interpolation*[3] | | | *225* | *502* | | *99* | | *87* | | *76* | | *68* | | *60* |
| **Model-zrc-2** | *EDT values through interpolation* | | | *250* | *557* | | *98* | | *86* | | *76* | | *68* | | *60* |
| **Model-zrc-3** | *EDT values through interpolation* | | | *275* | *613* | | *97* | | *85* | | *75* | | *67* | | *59* |
| **N17** | **821** | 0.0064 | 70.74 | 350 | 820 | 4.819E-13 | 91 | 2.135E-13 | 80 | 1.199E-13 | 72 | 6.544E-14 | 63 | 3.462E-14 | 55 |

*1 - eU estimated by iterating alpha-dose tht is equivalent to published value after radiaiton damage accumulation period of 3950 Myrs.*

*2 - EDT caluclated after Tremblay et al. (2014).*

*3 - EDT values for model zircons determined thorugh interpolation from published data for diffusivity, activation energy, mean diffusivity and eU*

**Supplementary Table S5:** Accumulation times required for zircon with eU = 10-300 ppm to reach the first percolation threshold of radiation damage based on alpha-dose values presented by Pidgeon et al. (2014) and Zhang et al. (????). Where accumulationtimes require >4400 Myrs to reach the threshold, calculations are for 4400 Myrs. Also given (right side of table).

| Zircon composition | Radiation damage accumulation time | | First percolation threshold (alpha-dose / gram) | | Accum. Radiation damage (alpha-dose / gram) for periods pertinent to lunar events | | |
|---|---|---|---|---|---|---|---|
| | | | Pidgeon et al. (2014) | Zhang et al. (????) | | | |
| eU (ppm | Myrs | Myrs | 2.2E+18 | 3.0E+18 | 3950 Myrs* | 3850 Myrs** | 100 Myrs*** |
| 300 | **1868** | **2380** | 2.2E+18 | 3.0E+18 | 6.7E+18 | 6.4E+18 | 9.8E+16 |
| 250 | **2160** | **2720** | 2.2E+18 | 3.0E+18 | 5.6E+18 | 5.3E+18 | 8.2E+16 |
| 200 | **2554** | **3155** | 2.2E+18 | 3.0E+18 | 4.5E+18 | 4.2E+18 | 6.6E+16 |
| 150 | **3107** | **3735** | 2.2E+18 | 3.0E+18 | 3.3E+18 | 3.2E+18 | 4.9E+16 |
| 105 | **3755** | **4350** | 2.2E+18 | 3.0E+18 | 2.4E+18 | 2.3E+18 | 3.4E+16 |
| 100 | **3925** | **4400** | 2.2E+18 | 2.8E+18 | 2.2E+18 | 2.1E+18 | 3.3E+16 |
| 75 | **4397** | **4400** | 2.2E+18 | 2.2E+18 | 1.7E+18 | 1.7E+18 | 2.5E+16 |
| 50 | **4400** | **4400** | 1.4E+18 | 1.4E+18 | 1.1E+18 | 1.1E+18 | 1.6E+16 |
| 30 | **4400** | **4400** | 8.4E+17 | 8.4E+17 | 6.7E+17 | 6.4E+17 | 9.8E+15 |
| 20 | **4400** | **4400** | 5.6E+17 | 5.6E+17 | 4.5E+17 | 4.2E+17 | 6.6E+15 |
| 10 | **4400** | **4400** | 2.8E+17 | 2.8E+17 | 2.2E+17 | 2.1E+17 | 3.3E+15 |

* Radiation damage dose assuming zircon grains did not experience a thermal event at 100 Ma
** Radiation damage dose of zircon grains prior to a thermal event at 100 Ma
*** Present day radiaition damage dose of zircon grains assuming complete annealing of all grains at 100 Ma

**Supplementary Table S6:** Summary of minimum temperatures required for annealing across all zircon compositions - predicted by HeFTy using the ZrDAAM kinetic model (Guenthner et al., 2013)

| Cooling duration @ 100 Ma | Partial annealing (T°C) | Complete annealing (T°C) | Max event T allowed |
|---|---|---|---|
| 100 kyrs | >200°C | >400°C | 220 |
| 10 kyrs | >225°C | >400°C | 255 |
| 2 kyrs | >250°C | >400°C | 275 |
| 1 yr | >375°C | >500°C | 400 |
| 1 mth | >400°C | >600°C | 500 |
| 1 day | >500°C | >700°C | 630 |
| 1 min | >600°C | >835°C | 930 |